\documentclass[
aps,
 amsmath,
 amssymb,
reprint,
]{revtex4-2}

\usepackage{graphicx}
\usepackage{dcolumn}
\usepackage{bm}
\usepackage{relsize}
\usepackage{subfig}
\usepackage{float}
\usepackage{upgreek}

\begin{document}


\title{Effect of junction critical current disorder in superconducting quantum interference filter arrays}  


\author{K.-H. M\"{u}ller}
\email{karl.muller@csiro.au}
\author{E. E. Mitchell}

\affiliation{CSIRO Manufacturing, PO Box 218, Lindfield, NSW 2070, Australia
}%

\date{\today}

\begin{abstract}
In this study, we investigated the performance of two 2D superconducting quantum interference filter (SQIF) arrays fabricated from YBCO thin films at a temperature of 77 K. Each array consisted of 6 Josephson junctions (JJs) in parallel and 167 in series. We conducted both experimental and theoretical analyses, measuring the arrays' voltage responses to an applied magnetic field and their voltage versus bias-current characteristics.
To properly model the planar array layouts, our theoretical model used the stream function approach and also included the Johnson noise in the JJs. The model further divides the superconducting current density of the arrays into its Meissner current, circulating current, and bias current parts for practicality. Since the fabrication process of YBCO thin films cannot produce identical JJ critical currents, we assumed a log-normal distribution to model the JJ critical current disorder. Our model predictions, with a JJ critical current spread of 50\%, agreed well with our experimental data.
Using our model, we were able to study the dependence of the voltage modulation depth on critical current disorder and London penetration depth. We also analyzed the observed reflection asymmetries of the voltage versus magnetic field characteristics, which might provide insight into the degree of critical current disorder.
Overall, our findings suggest that the use of YBCO thin films in SQIF arrays is promising, despite the critical current disorder inherent in their fabrication process. Our study highlights the importance of theoretical modeling in understanding the performance of superconducting devices and provides insights that could inform the design of future SQIF arrays.
\end{abstract}

\maketitle

\section{\label{sec:level1}Introduction}

When fabricating 2D superconducting quantum interference filter (SQIF)  arrays (or superconducting quantum interference (SQUID) arrays) from YBCO thin films, the critical currents $I_{ck}$ and electric resistances $R_k$ of the Josephson junctions (JJs) exhibit significant variability and thus have wide distributions. This  $I_c$-$R$-disorder in the arrays is thought to result in a decrease in the magnetic-field-to-voltage transfer function of the array, which is an undesirable outcome as the goal is typically to achieve a transfer function that is as large as possible for optimal performance in applications. How detrimental such an $I_c$-$R$-disorder is in real devices, is still unknown. This is in contrast to low-temperature superconducting thin-film arrays, where the $I_{ck}$ and $R_k$ values can be well controlled.

In this paper, we aim to provide a quantitative analysis of the impact of the $I_c$-$R$-disorder on the performance of two specific YBCO 2D SQIF arrays that we have recently fabricated and measured. Our discussion will focus on the effects of $I_c$-$R$-disorder in these particular arrays, while the broader impact of $I_c$-$R$-disorder on 2D SQIF (or SQUID) arrays of varying dimensions and with different screening and thermal noise parameters will be addressed in a forthcoming publication.

Previously, we extensively modeled a structurally wide, thin-film 1D parallel SQUID array with center current biasing and used a brute-force approach to calculate the voltage-to-magnetic-field characteristics, as described in reference \cite{MUL21}. In the present paper, we use an improved version of our previous model, where we separate the superconducting current density into its Meissner current, circulating loop currents, and bias-current components. This separation allows us to save significant computational time and accurately calculate larger 2D arrays with strong $I_c$-$R$-disorder.

Previous theoretical studies on 1D and 2D SQUID \cite{CYB12, DAL14, TAY16, MIT19}, SQIF \cite{OPP01} and JJ \cite{REI94} arrays have often not considered the effects of thermal noise. Recent research has shown that properly accounting for thermal noise is essential for accurately calculating the voltage-to-magnetic-field characteristics and transfer function of these arrays \cite{GAL22, GAL22b, PEG23}. Only a few studies so far have investigated the consequences of $I_c$-$R$-disorder in arrays \cite{BER15, MIT19, CRE21}, but direct comparisons with experimental data from genuine arrays with wide thin-film structures are lacking.

In this paper, we begin by discussing the layout, fabrication, and measurement setup of our 2D SQIF arrays in Section II. We then present our experimental results in Section III. In Section IV, we outline our theoretical model, including the JJ phase dynamics and the stream function approach. In Section V, we compare our experimental data with our model calculations and discuss the insights gained from our detailed theoretical calculations with emphasis on the $I_c$-$R$-disorder. We summarize our findings in Section VI. Appendices are included to provide guidance on how to obtain the effective hole areas and the geometric and kinetic inductances.

\section{2D SQIF array layout, fabrication and measurement setup}

Two different SQIF arrays were fabricated lithographically from a thin-film of YBCO grown by e-beam evaporation on a 1 cm$^2$ MgO substrate. The thickness of the film was $d$ =113 nm and its superconducting transition temperature was $T_c = 85.9$ K. Before film deposition, steps were etched into the MgO surface (see Fig. 1) using a well established technique based on argon milling \cite{FOL99,MIT10}. During YBCO thin-film deposition, grain boundaries form along the edges of the sharp MgO steps, producing JJs. Films were then lithographically patterned into two different 2D SQIF arrays referred to in the following as arrays S1 and S2, which are displayed in Fig. 1. 
\par
\begin{figure}[h]
\begin{center}
\hspace*{-2mm}
\includegraphics[width=0.40\textwidth]{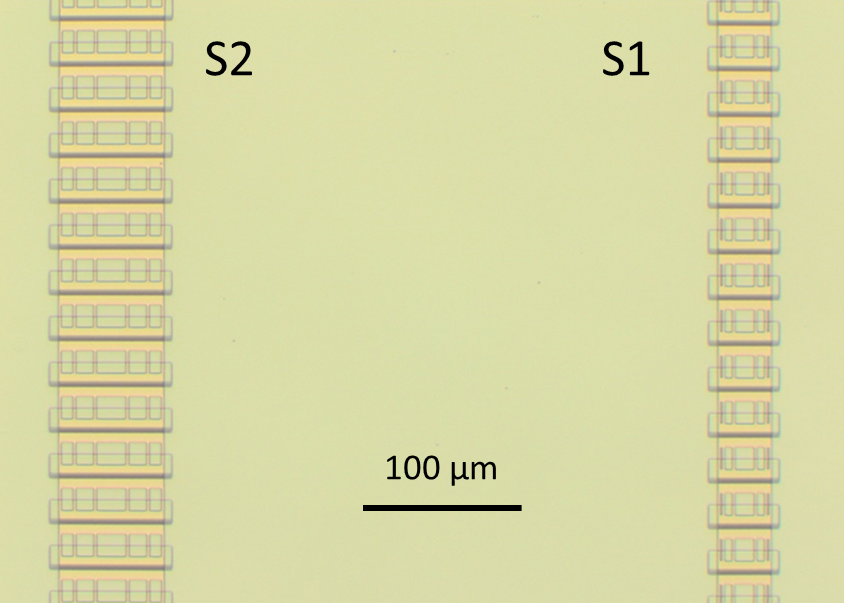}
\label{default}
\hspace*{-3mm}
\includegraphics[width=0.40\textwidth]{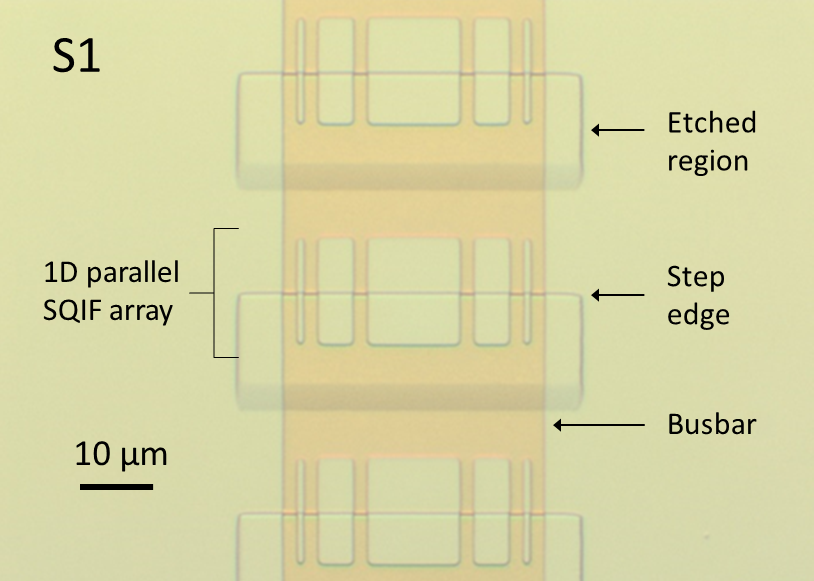}
\caption{2D SQIF arrays S1 and S2. Etched rectangular regions in the substrate are visible. Each of these regions has a pronounced sharp step edge at the top to promote grain boundary thin-film growth and a very gradual step on the bottom to avoid grain boundary formation in the busbar region.}
\label{default}
\end{center}
\end{figure}
Each array consists of $N_s = 167$ identical 1D parallel SQIF arrays (the rows) which are connected in series via superconducting YBCO thin-film busbars. Each 1D SQIF array has $N_p$ = 6 JJs in parallel. Table I lists the design parameters of S1 and S2, {\it{i.e.}} the SQIF hole height, the hole widths (left to right), the busbar height, and the number $N_s$ of identical 1D parallel SQIF arrays connected in series. The tracks leading to the JJ's have the same width as the JJs themselves (Table I). Both arrays have geometric left-right symmetry. Because of the difficulty controlling grain boundary growth, the values of the critical currents $I_{ck}$ and resistances $R_k$ ($k$ = 1 to $N_s N_p$) of the JJs can vary widely \cite{LAM14}. This disorder in $I_{ck}$ and $R_k$ negatively affects the applied magnetic field to voltage transduction. In a 2D array, it is not possible to measure the $I_{ck}$ and $R_k$ values individually and only consequences of the $I_c$-$R$-disorder can be detected.\\

\begin{table} [! h]
\begin{center}
\caption{SQIF array design parameters}
\label{tab:table1}
\begin{tabular}{|c || c | c | c | c | c |} 
 \hline
  \# &hole height&hole widths  L-R&busbar height&JJ width&$N_s$ 
\\
 & [ $\mu$m ] & [ $\mu$m ] & [ $\mu$m ] & [$\mu$m] &
\\ 
\hline\hline
 S1 & 15 & 1, 5, 13, 5, 1 & 16 & 2 & 167 \\ 
\hline
S2 & 15 & 8, 12, 20, 12, 8 & 16 & 2 & 167 \\
\hline
\end{tabular}
\end{center}
\end{table}

For measurements, the arrays were placed on a measurement probe and dipped into a dewar of liquid nitrogen and zero field cooled from room temperature down to 77 K. To screen out  the earth's magnetic field, the dewar was surrounded by five layers of mu-metal shielding. The standard four-terminal method was used to measure the DC voltages $V$ appearing between the ends (between top and bottom) of the 2D SQIF arrays, while a bias current $I_b^{tot}$ was injected at the top and exited the array at the bottom. A solenoid surrounding the probe enabled us to generate a homogeneous perpendicular applied magnetic field $B_a$ at the array. Both $V(B_a)$ and $V(I_b^{tot})$ of S1 and S2 were measured at 77 K.

\section{\label{sec:level1}Experimental results}
Figure 2 shows the experimental results of voltage $V$ versus the applied magnetic field $B_a$ for arrays S1 and S2, where $B_a$ was varied from $ - 50 \, \mu\text{T}$ to $ + 50 \, \mu\text{T}$. 

\begin{figure}[h]
\begin{center}
\hspace*{-7mm}
\includegraphics[width=0.60\textwidth]{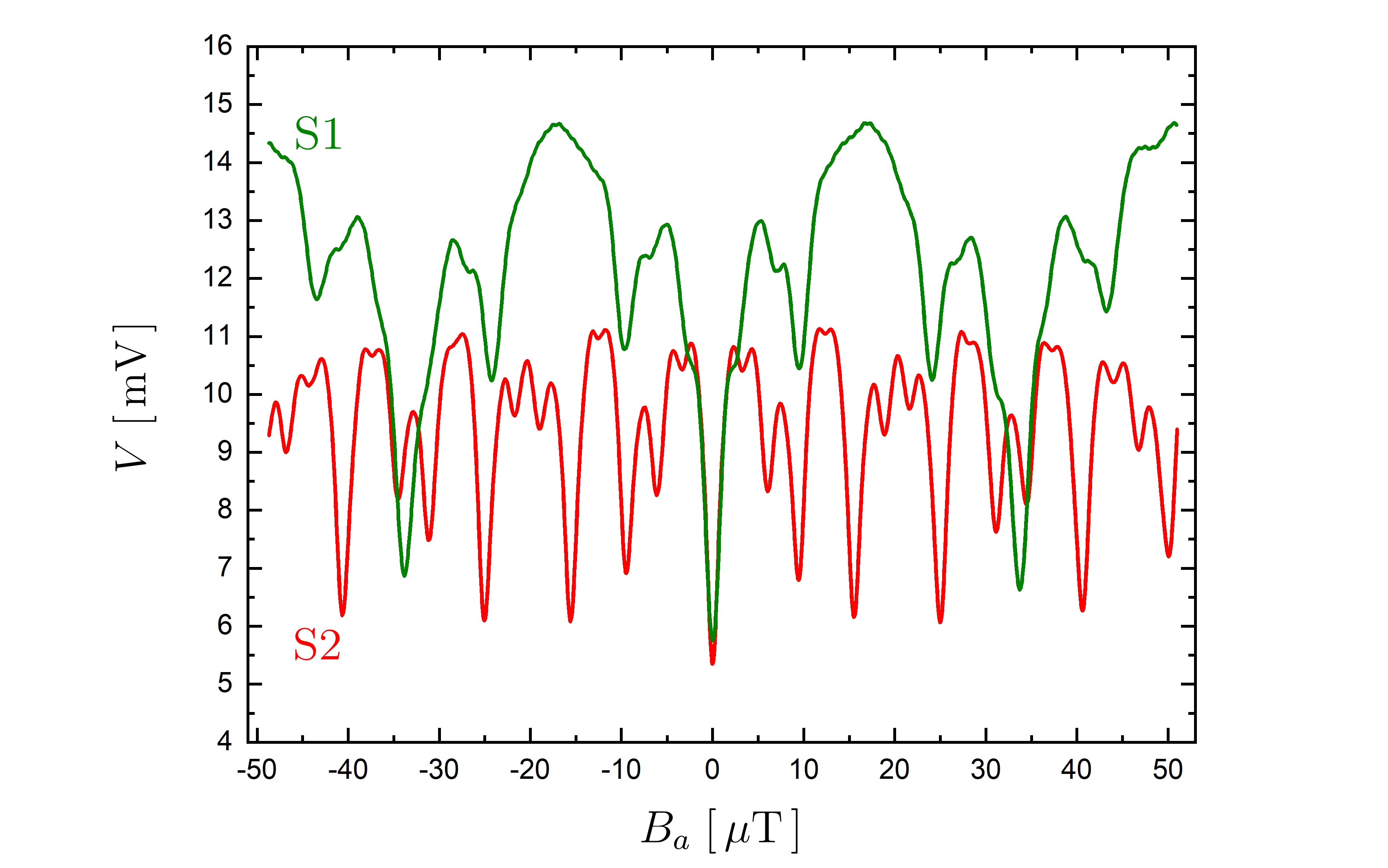}
\caption{Experimental data $V(B_a)$ of arrays S1 and S2 for optimal bias current $I_{b, opt}^{tot}$ (see Table II).}
\label{default}
\end{center}
\end{figure}

The bias currents $I_b^{tot}$ for S1 and S2 were chosen to optimize the voltage modulation depths $\Delta V$. $\Delta V$ can be used as a measure of the maximum transfer function  $dV / dB_a |_{max}$ of the center dip \cite{GAL22}. The values of the optimal bias current $I_{b, opt}^{tot}$ and $\Delta V$ are given in Table II. Both $V(B_a)$ curves are nearly reflection symmetric with respect to the center dip. Because the rectangular holes in S2 are overall larger than those in S1, the $V(B_a)$ characteristics of array S2 shows more oscillations in Fig. 2 than S1. To understand the origin of the complex shapes of these two $V(B_a)$ characteristics, we have performed theoretical modeling which is outlined in detail below.

\begin{table} [! h]
\begin{center}
\caption{SQIF array electrical parameters}
\label{tab:table1}
\begin{tabular}{|c || c | c | c | c |} 
 \hline
 \#  & $I_{b, opt}^{tot}$ [ $\mu$A ]  & $\Delta V$ [mV] & $R^{tot}$ [ $\Omega$ ] 
\\ 
\hline\hline
S1 & 76.474 & 8.52 & 406.0  \\
\hline
S2 & 78.215 & 5.76 &  399.1  \\ 
\hline
\end{tabular}
\end{center}
\end{table}

Figure 3 displays the results of our $V(I_b^{tot})$ characteristics measurements in zero magnetic field for S1 and S2. The differential resistance at large $I_b^{tot}$ defines the total resistance $R^{tot}$ of the array which is given in Table II. Our theoretical modeling results in Section V will give an explanation for the shape of $V(I_b^{tot})$.

\begin{figure}[h]
\begin{center}
\hspace*{-6mm}
\includegraphics[width=0.57\textwidth]{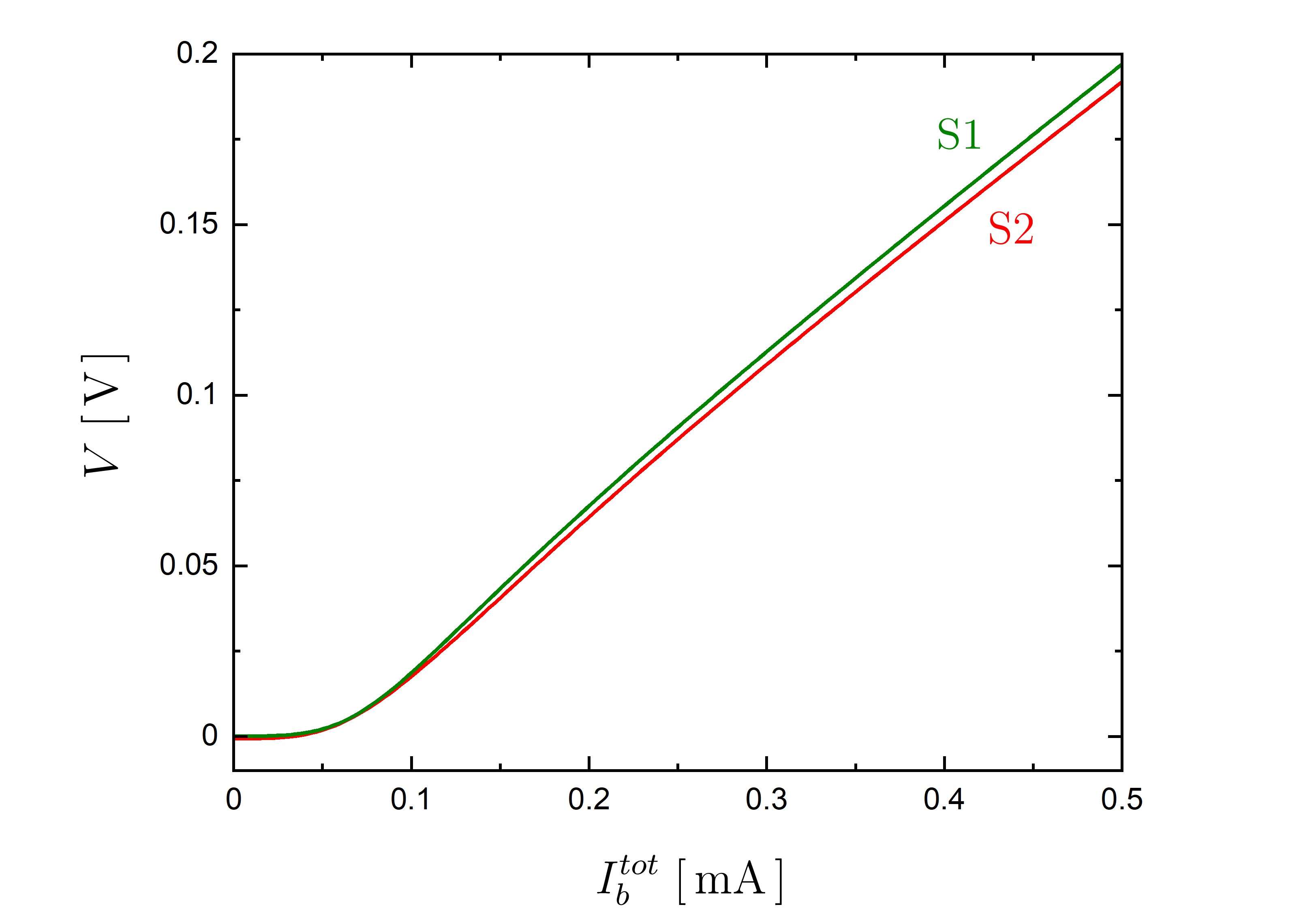}
\label{default}
\caption{Experimental data of voltage $V$ versus bias current $I_b^{tot}$ of S1 and S2 in zero applied magnetic field.}
\label{default}
\end{center}
\end{figure}

Section IV below will outline the theoretical model that we use to describe 2D SQIF (and SQUID) arrays. We will see how our model allows us to extract from our experimental data information about  the JJ $I_c$-$R$-disorder in our fabricated arrays. Furthermore, our model predicts how the voltage modulation depth $\Delta V$, which can be used as a measure for the maximum transfer function $dV / dB_a |_{max}$, decreases with increasing London penetration depth $\lambda$ and increasing JJ $I_c$-$R$-disorder.
Our model also elucidates the role of the row-row mutual inductive coupling in 2D SQIF arrays with busbars.

\section{\label{sec:level1}Theoretical model}

\subsection{\label{sec:level2}JJ phase dynamics}

Our fabricated step-edge JJs have negligibly small capacitances and behave like short JJs as the JJ width is much less than their Josephson penetration depths at 77 K. Therefore, we use the resistively shunted junction model (RSJ model) to describe our 2D SQIF arrays. The current $I_k$ flowing through the $k^{th}$ JJ (Fig. 4), where $k$ = 1 to $N_p N_s$ (numbered from left to right through the rows from top to bottom), is the sum of three different currents: (i) the current through the intrinsic shunt resistance $R_k$, (ii) the Josephson current $I_{ck} \sin \varphi_k$, and (iii) the noise current $I_{Nk}$. Here $I_{ck}$ is the JJ critical current and $\varphi_k$ is the gauge invariant phase difference.  The noise current $I_{Nk}$ is due to the Johnson noise voltage which appears across the JJ intrinsic shunt resistors $R_k$ at finite temperature $T$.

\begin{figure}[h]
\begin{center}
\hspace*{+2mm}
\includegraphics[width=0.45\textwidth]{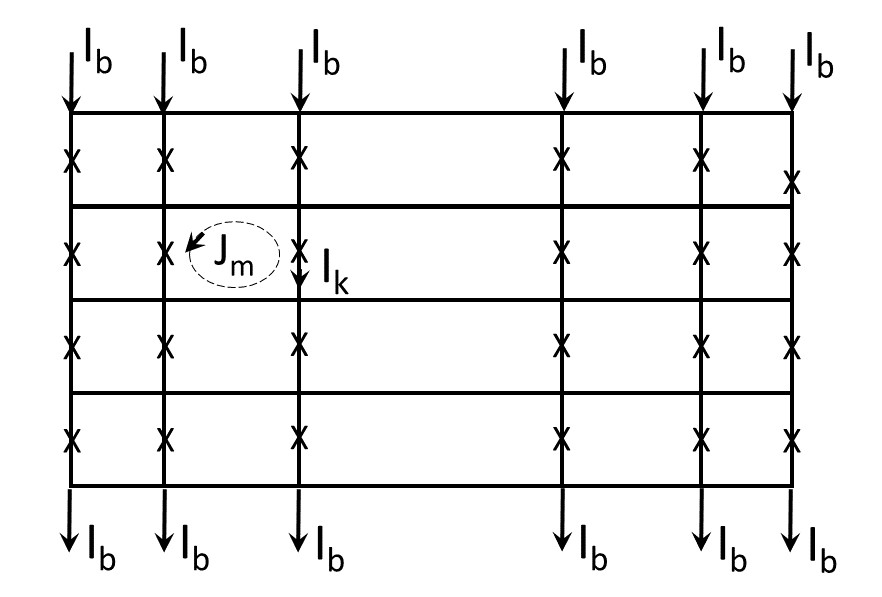}
\caption{Current flow schematic for a 2D SQIF array with uniform bias current injection. The crosses mark the resistively shunted Josephson junctions. Here, $N_p = 6$ and $N_s = 4$. For simplicity, only one circulating current $J_m$ and one current $I_k$ flowing through the $k^{th}$ JJ are indicated.}
\label{default}
\end{center}
\end{figure}

Employing the Josephson equation for the voltage across the $k^{th}$ JJ, one obtains
\begin{equation}
\frac{\Phi_0}{2 \pi R_k} \frac{d \varphi_k(t)}{dt} + I_{ck} \sin \varphi_k(t) + I_{Nk}(t) = I_k(t) \; .
\end{equation}
\\
Here $\Phi_0$ is the flux quantum and $t$ the time.
Introducing the dimensionless time $\tau = 2 \pi R I_c t / \Phi_0$ where $R$ is the average JJ resistance and $I_c$ the average critical JJ current, one can write Eq. (1) in vector notation as
\begin{equation}
\hat{\xi}^{-1} \frac{d \vec{\varphi}(\tau)}{d \tau} + \hat{\eta} \; \overrightarrow{\sin \varphi(\tau)} + \vec{i}_N (\tau) = \vec{i} (\tau) \; .
\end{equation}
Here $\hat{\xi}$ and $\hat{\eta}$ are $N_s Np \times N_s N_p$ diagonal matrices with diagonal elements $\hat{\xi}_{kk} = R_k / R$ and $\hat{\eta}_{kk} = I_{ck} / I_c$, respectively. The components of $\vec{\varphi}$ are $\varphi_k$ and those of  $\overrightarrow{\sin \varphi}$ are $\sin \varphi_k$, while the components of $\vec{i}_N$ are $I_{Nk} / I_c$ and those of $\vec{i}$ are $I_k / I_c$.

From Kirchhoff's law one obtains for a 2D array with uniform bias current injection as indicated schematically in Fig. 4
\begin{equation}
\vec{i}(\tau)= \hat{K} \, \vec{J} (\tau)/ I_c + \vec{i}_b \; .
\end{equation}
Here $\vec{J}$ is the circulating current vector with components $J_m$ flowing around hole $m$ (Fig. 4), where $m$ = 1 to $N_s (N_p -1)$, numbered from left to right through the rows from top to bottom. The transport bias current vector $\vec{i}_b$ is $N_s N_p$ dimensional with identical components $I_b / I_c$. The total bias current is $I_b^{tot} = N_p \, I_b$ where the time independent bias current is uniformly injected from the top as shown in Fig 4. Because in our arrays $N_s >> N_p$, any effects arising from the top and bottom bias current leads are neglected. The matrix $\hat{K}$ in Eq. (3) is an $N_s N_p \times N_s (N_p - 1)$ matrix of the form 
\begin{equation}
\hat{K} = \hat{I}_{N_s \times N_s}  \otimes \hat{\kappa} \; ,
\end{equation}
where $\otimes$ is a tensor product with $\hat{I}_{N_s \times N_s}$ an $N_s \times N_s$ identity matrix and $\hat{\kappa}$ an $N_p \times (N_p -1)$ matrix with
elements $(\hat{\kappa})_{ij} = \delta_{ij} - \delta_{i-1, j}$. Here $\delta_{ij}$  is the Kronecker $\delta$. 

Using the second Ginzburg-Landau equation one can relate neighboring JJ phases within each row via the expression
\begin{equation}
\frac{\Phi_0}{2 \pi} (\varphi_{k+1}  - \varphi_k) = \Phi_m+ \mu_0 \, \lambda^2 \oint_{C_m} \vec{j} \, d\vec{l} \; .
\end{equation}
\\
Here $\lambda$ is the London penetration depth.
The index $m$ is the index for the circulating currents $J_m$ and for the holes. (The index $m$  is related to the JJ index $k$ via the relation $k = m + \text{Int}\{m/(N_p-1) - \epsilon\}$ where $\text{Int}\{ \}$ is the integer part of its argument and $\epsilon$ is an infinitesimally small positive real number.) $\Phi_m$ in Eq. (5) is the magnetic flux through hole $m$, and $\vec{j}$ is the super-current density in the array. The line integration in Eq. (5) is performed counterclockwise along the contour $C_m$ which runs along the edges of the hole, and $\mu_0$ is the permeability of free space.
\\
\par
The magnetic flux $\Phi_m$ in Eq. (5) is made up of four contributions,

\begin{equation}
\Phi_m= \, \Phi^{(\text{app})}_m+ \Phi^{(\text{Mei})}_m + \Phi^{(J)}_m + \Phi^{(I_b)}_m \; .
\end{equation}
\\
Here, $\Phi^{(\text{app)}}_m$ is the externally applied flux through hole $m$, with $\Phi^{(\text{app})}_m = A_m B_a$, where $A_m$ is the hole area and $B_a$ the applied homogeneous perpendicular magnetic field. $\Phi_m^{\text{(Mei)}}$ is the flux through hole $m$ created by all the Meissner shielding currents induced in response to the applied field $B_a$.
$\Phi_m^{(J)}$ is the flux produced in hole $m$ by all the circulating currents $\vec{J}$, and $\Phi_m^{(I_b)}$ is the flux through hole $m$ produced by the transport bias currents flowing vertically through all the JJ's from the top to the bottom (Eq. (3) and Fig. 4). 
\par
In a similar way, the supercurrent density $\vec{j}$ in Eq. (5) is made up of three parts, 
\\
\begin{equation}
\vec{j}\,  = \, \vec{j}^{(Mei)} + \vec{j}^{(J)} + \vec{j}^{(I_b)}  \; .
\end{equation}
\\
Here $\vec{j}^{(\text{Mei})}$ is
the Meissner shielding current density, $\vec{j}^{(J)}$ the current density from all the circulating currents, and $\vec{j}^{(I_b)}$ the transport bias current density in the array. The bias current $I_b$ not only flows through the leads (Fig. 4), but also through each JJ according to Eq. (3).
\\
\par
By using Eqs. (6) and (7), one can rewrite Eq. (5) as
\begin{equation}
\frac{\Phi_0}{2 \pi} (\varphi_{k+1}  - \varphi_k) = \Phi_{1,m} + \Phi_{2,m} + \Phi_{3,m}\; ,
\end{equation}
\\
where the fluxoids $\Phi_{1,m}$,  $\Phi_{2,m}$ and $\Phi_{3,m}$ are defined as
\\
\begin{equation}
\Phi_{1,m} = \Phi_m^{(\text{app})} + \Phi_m^{(\text{Mei})} +  \mu_0 \lambda^2 \oint_{C_m} \vec{j}^{(\text{Mei})} \, d\vec{l} \; ,
\end{equation}
\begin{equation}
\Phi_{2,m} = \Phi_m^{(\text{J})} +  \mu_0 \lambda^2 \oint_{C_m} \vec{j}^{(\text{J})} \, d\vec{l}
\end{equation}
and
\begin{equation}
\Phi_{3,m} = \Phi_m^{(I_b)} +  \mu_0 \lambda^2 \oint_{C_m} \vec{j}^{(I_b)} \, d\vec{l} \; .
\end{equation}
\\
The fluxoid $\Phi_{1,m}$ in Eq. (9) can be written as
\\
\begin{equation}
\Phi_{1,m} = B_a A_m^{\text{eff}} \; ,
\end{equation}
\\
where $A_m^{\text{eff}}$ is the effective hole area of hole $m$, with
\\
\begin{equation}
A_m^{\text{eff}} = \frac{B_a A_m + \Phi^{(\text{Mei})} + \mu_0 \lambda^2 \oint_{C_m} \vec{j}^{(\text{Mei})}  \, d\vec{l} } {B_a} \, .
\end{equation}
\\
The effective hole area $A_m^{\text{eff}}$ does not depend on $B_a$, because both $\Phi^{(\text{Mei})}$ and  $\vec{j}^{(\text{Mei})}$ are proportional to $B_a$. How we 
determine $A_m^{\text{eff}}$ is outlined further below.
\\
\par
Furthermore, the fluxoid $\Phi_{2,m}$ in Eq. (10) can be written as
\\
\begin{equation}
\Phi_{2,m} = (\hat{L} \,\vec{J})_m \; ,
\end{equation}
\\
where $\hat{L}$ is the $N_s (N_p - 1) \times N_s (N_p - 1)$ inductance square matrix of the array, with each matrix element the sum of a geometric and kinetic term.
How we determine the inductance matrix $\hat{L}$ is outlined further below.
\\
\par
And, the fluxoid $\Phi_{3,m}$ in Eq. (11) can be written as
\begin{equation}
\Phi_{3,m} = (\vec{L}^{I_b})_m \, I_b \; ,
\end{equation}
where we name $\vec{L}^{I_b}$ the bias current inductance vector which has $N_s (N_p - 1)$ components. Each component is the sum of a geometric and kinetic term.
How we determine the bias current inductance vector $\vec{L}^{I_b}$ is outlined further below.
\\
\par
Finally, using Eqs. (2), (3), and (8) - (15), we derive the following system of nonlinear stochastic differential equation of first order for the time evolution of the JJ phase difference vector $\vec{\varphi}(\tau)$
\\
\[
\hat{\xi}^{-1} \frac{d \vec{\varphi}(\tau)}{d \tau} + \hat{\eta} \; \overrightarrow{\sin \varphi (\tau)} + \vec{i}_N(\tau)  =
\]
\begin{equation}
 \vec{i_b} \, + \, \frac{\hat{K} \hat{L}^{-1}} {I_c} \,  [ \; \frac{\Phi_0}{2 \pi} \hat{D} \vec{\varphi}(\tau)
- B_a \vec{A}^{\text{eff}}  - \vec{L}^{I_b} \, I_b \;] \; .
\end{equation}
\\
Here $\hat{D}$ is an $N_s N_p \times N_s (N_p - 1)$ matrix of the form
\\
\begin{equation}
\hat{D} = \hat{I}_{N_s \times N_s}  \otimes \hat{\delta} \; ,
\end{equation}
\\
where $\hat{\delta}$ is an $(N_p -1) \times N_p$ matrix with elements $(\hat{\delta})_{ij}  = -\delta_{ij} + \delta_{i,j-1}$. The differential equation Eq. (16) is the key equation of this paper.
\\
\par
The set of dynamic equations for the JJ phase differences $\varphi_k$ (Eq. (16)) can also be written with the help of a so-called washboard potential $U(\{\varphi_k\})$ as \cite{REI94} 
\begin{equation}
\frac{\Phi_0^2}{4 \pi^2 R_k} \frac{d \varphi_k}{d t} = - \frac{\partial U(\{\varphi_k\})}{\partial \varphi_k} \; ,
\end{equation}
where
\begin{equation}
U(\{\varphi_k\}) = - \sum_{k=1}^{N_s N_p} [ \, \frac{\Phi_0}{2 \pi} I_{ck} \cos \varphi_k + \frac{\Phi_0}{2 \pi} (I_b  - I_{Nk} ) \varphi_k  \,] 
\end{equation}
\[ + \, \frac{1}{2} \sum_{m=1, m'=1}^{N_s (N_p -1)} M_m  \, \hat{L}^{-1}_{m, m'} \, M_{m'} \; ,
\]
with
\begin{equation}
M_m = \frac{\Phi_0}{2 \pi} ( \varphi_{k+1} - \varphi_k) - A_m^{\text{eff}} B_a - L_m^{I_b} I_b \; ,
\end{equation}
where $k$ is a function of $m$ as stated above.

The potential $U$ can be obtained from energy considerations or conjectured directly from Eq. (16). Evaluating the right hand side of Eq. (18) leads to Eq. (16). 
\\
\par
The time averaged voltage $V$ of the SQIF array, measured between its top and bottom ends, is given by 
\begin{equation}
V = R  \, I_c \lim_{{\tau} \rightarrow \infty} \, \frac{1}{\tau} \, \frac{1}{N_p} \sum_{k=1}^{N_s N_p}  [ \, \varphi_k(\tau) - \varphi_k(0) \, ] \; .
\end{equation}
\\
\par
The noise current vector $\vec i_N(\tau)$, which appears in Eq. (16), was implemented using uncorrelated Gaussian random variables with the mean-square deviations $2 \, \Gamma_k / \Delta \tau$. Here $\Delta \tau$ is the time-step interval used in the finite difference method used to numerically solves Eq. (16), and $\Gamma_k$ is the noise strength defined as
\begin{equation}
\Gamma_k = \frac{R} {R_k} \, \frac{2 \pi k_B T}{I_c \Phi_0} \, ,
\end{equation}
where $k_B$ is the Boltzmann constant and $T$ the operating temperature of the array.
\\
\par
\subsection{\label{sec:level2}Stream function approach}

In this section we show how, for superconducting thin-film arrays with extended busbar structures (Fig. 1), one can use the stream function approach \cite{KHA01, KHA05, BRA05, KHA10, MUL21, BIS22} to calculate the effective hole areas $\vec{A}^{\text{eff}}$ (Eq. (13)), the array inductance matrix $\hat{L}$ (Eq. (14)), and the bias current inductance vector $\vec{L}^{I_b}$ (Eq. (15)). These three physical quantities are needed in order to numerically solve the set of differential equations in Eq. (16) and to obtain the time evolution of the JJ phase differences $\varphi_k(\tau)$. 
\\
\par
For a superconducting thin-film of thickness $d$ in the $xy$-plane, the super-current density $\vec{j}(x,y,z)$ becomes independent of the perpendicular direction $z$ if $\lambda > d$ \cite{CLE05,BRA05}. In our case we have $d = 0.113 \, \mu$m and $\lambda = 0.42 \, \mu$m (as shown later) and thus $\lambda > d$. In this case one can introduce a 2D stream function $g(x,y)$ which is defined via the $x$ and $y$ components of $\vec{j}$ as
\begin{equation}
j_x  = \frac{1}{d} \frac{\partial g}{\partial y} \; \; \text{and} \; \; j_y = - \frac{1}{d} \frac{\partial g}{\partial x} \; .
\end{equation}
\\
Using Eq. (23) together with the second London equation and Biot-Savart's law in 2D, one obtains by applying partial integration a second-order linear Fredholm intregro-differential equation for the stream function $g(x,y)$ of the form

\[
\frac{\lambda^2}{d}  \, ( \frac{\partial^2}{\partial x^2} + \frac{\partial^2}{\partial y^2} ) \, g(x,y)  \; +
\]
\begin{equation} 
\frac{1}{4 \pi} \int_{\Omega} Q(x,y,x',y') \; g(x',y') \; dx' dy' - f_s(x,y) \, = \, \frac{B_a}{\mu_0}  \; ,
\end{equation}
where $B_a$ is the applied magnetic field pointing in $z$ direction and
 \begin{equation}
f_s(x,y) \; =
\end{equation}
\[
\frac{1}{4 \pi} \oint_{\partial \Omega} \frac{g(x',y')}{\sqrt{(x-x')^2 + (y-y')^2}^{\;3} }\, 
 \left( {\begin{array}{cc}
   x-x' \\
   y-y' \\
  \end{array} } \right)
\vec{n} \; dl' \, .
\]
\\
In Eq. (24), $\Omega$ is the integration domain, {\it{i.e.}} the superconducting thin film area, and $\partial \Omega$ is the domain boundary which includes the edges of the holes. The vector $\vec{n}$ in Eq. (25) is a normal vector in the $xy$-plane, perpendicular to the boundary, and $\vec{n}$ points outwards, away from the domain $\Omega$. In Eq. (25), $d l'$ is a line element.
\\
\par

The kernel $Q$ in Eq. (24) can be written as \cite{MUL21}
\begin{equation}
Q(x,y,x',y') = 
\end{equation}
\[
\frac{1}{\Delta x \Delta y} \, \Bigg[ \, \Big[ \frac{\sqrt{\bar{x}^2 + \bar{y}^2}}{\bar{x} \, \bar{y}} \Big]_{x'-x -\Delta{x}/2}^{x'-x+\Delta{x}/2} (\bar{x}) \Bigg]_{y'-y -\Delta{y}/2}^{y'-y+\Delta{y}/2} (\bar{y}) \; \;  ,
\] \\
with the definition
\begin{equation}
\Bigg[ \, \Big[ f(x,y) \Big]_{s_1}^{s_2} (x) \Bigg]_{s_3}^{s_4} (y) \, = 
\end{equation}
\[
f(s_2,s_4) - f(s_1,s_4) - f(s_2,s_3)+f(s_1,s_3) \; ,
\] \\
where $\Delta x$ and $\Delta y$ are the grid spacings used when solving solving Eq. (24) numerically. We used $\Delta x$ = $\Delta y$.
\\
\par
Having determined the stream function $g(x,y)$ allows us to calculate the magnetic fluxes $\Phi_m$ of the holes $m$ as well as the $\oint_{\partial \Omega_m} \vec{j} \, d\vec{l}$ terms. One obtains
\begin{equation}
\Phi_m = \mu_0 \int_{\Omega_m} h(x,y) \; dx \, dy \;  +  \, \Phi_m^{(\text{app})},
\end{equation}
where $h(x,y)$ is defined as
\begin{equation}
h(x,y) = f_s(x,y) \, -  \, \frac{1}{4 \pi} \int_{\Omega} \, Q(x,y,x',y') \; g(x',y') \; dx' dy' \; ,
\end{equation}
and
\begin{equation}
\oint_{\partial \Omega_m} \vec{j} \, d\vec{l} \, = \, \frac{1}{d} \, \oint_{\partial \Omega_m} \, (\, \frac{\partial g(x,y)}{\partial y} \, dx \, - \, \frac{\partial g(x,y)}{\partial x} \, dy \, ) \; .
 \end{equation}
\\
\par

By choosing different boundary conditions for the stream function $g(x,y)$, one can separately address the three different current densities in Eq. (7), {\it{i.e.}} $j^{(Mei)}$, ${j^{(J)}}$ and ${j^{(I_b)}}$, and then obtain  $\vec{A}^{\text{eff}}$, $\hat{L}$ and $\vec{L}^{I_b}$ from Eqs. (13) - (15). Further details about how to obtain $\vec{A}^{\text{eff}}$, $\hat{L}$ and $\vec{L}^{I_b}$ are given in Appendices A, B and C.
\\
\par

\section{\label{sec:level1}Model versus experimental results}

There are $2 N_s N_p +1$ unknown parameters in our model. These unknown parameters are the $N_s N_p = 167 \times 6 = 1002$ values for the JJ critical currents $I_{ck}$, the $N_s N_p = 1002$ values for the JJ resistances $R_k$, and the London penetration depth $\lambda$. Experimentally it has been found that in YBCO JJs, $I_{ck}$ and $R_k$ are approximately anti-correlated according to the empirical law $I_{ck} R_k \sim j_{ck}^{1/2}$, where $j_{ck}$ is the JJ critical current density \cite{GRO97, MIT10}. By exploiting this empirical law, we obtain for the diagonal matrix $\hat{\xi}$ in Eq. (16) the matrix elements $\hat{\xi}_{kk} =  \hat{\eta}_{kk}^{-1/2}$. This reduces the number of unknown model parameters by almost a factor of 2, from $2 \, N_s N_p + 1$ to $N_s N_p + 1$. 

The critical currents of YBCO step-edge junctions have a spread in values and previous measurements have found wide $I_{c}$-distributions of approximately Gaussian or even log-normal \cite{LAM14} shape. Therefore, when solving Eq. (16), we treat the $N_s N_p$ matrix elements $\hat{\eta}_{kk} = I_{ck} / I_c$ as log-normal random variables. A log-normal distribution $p(\eta)$ with mean value 1 and standard deviation $\sigma$ has the form 
\begin{equation}
p(\eta) = \frac{1}{\eta \, \gamma \, \sqrt{2 \pi}} \, \exp \frac{-(\ln \eta - \mu)^2}{2 \, \gamma^2} \; .
\end{equation}
Here $\mu = - \, \gamma^2 / 2$ and $\gamma = \sqrt{ \ln ( 1 + \sigma^2)}$. A log-normal distribution is similar to a Gaussian distribution for small $\sigma$. In our modeling, we cannot use Gaussian random variables for $\hat{\eta}_{kk}$ because they would lead to unphysical negative $I_{ck} / Ic$ values, particularly for large $\sigma$. In the following we call the standard deviation $\sigma$ the $I_c$-spread.

The number of unknown parameters is reduced slightly further by the relation between the measured total resistance $R^{tot}$ (Table II) and the unknown average JJ resistance $R$. This relation is given by 
\begin{equation}
R = \frac{R^{tot} }{\sum_{i=1}^{N_s} \, (\sum_{j=1}^{Np} \hat{\eta}_{kk \leftarrow ij}^{1/2})^{-1}} \; \, .
\end{equation}
Here the symbol $kk \leftarrow ij$ means that the $ij$ indices (where $i = 1$ to $N_s$ is the row index and $j = 1$ to $N_p$, the column index) are mapped onto the $k$ index which runs from 1 to $N_s N_p$. For arrays with sufficiently large $N_s N_p$, due to the introduction of $p(\eta)$ and Eq. (32), one no longer is dealing with $N_s N_p +1$ parameters but instead one only has to deal with three parameters, which are $I_c$, $\lambda$ and $\sigma$. Because S1 and S2 were produced from the same YBCO thin film on the same substrate, one can expect $\lambda$ and $\sigma$ to be the same 
for S1 and S2. But because of the finite values $N_p = 6$ and $N_s =167$, one can expect a slight difference between the average $I_c$ values of array S1 and S2.
The values of $I_c$, $\lambda$ and $\sigma$ for S1 and S2, that were found to best model our experimental data, are listed in Table III. 
These values were found by probing many different $I_c$, $\lambda$ and $\sigma$ combinations and comparing by eye the calculated $V(B_a)$ and $V(I_b^{tot})$ curves with the experimental data curves. The search for the $I_c$, $\lambda$ and $\sigma$ parameters was made easier by exploiting the facts that the $V(I_b^{tot})$ curves are very sensitive to $I_c$ and that decreasing $\lambda$ shifts $V(B_a)$ upwards and increases the voltage modulation depth $\Delta V$, whereas increasing $\sigma$ only decreases $\Delta V$. Because of the $I_c$-disorder and the finite size of the arrays, the parameters $I_c$, $\lambda$ and $\sigma$ can only be determined with our procedure with some uncertainties which we estimated to be about 2\%, 5\% and 10\%, respectively.
Previously, in DC-SQUIDs \cite{KEE21} that were made from similar YBCO thin films, a value of $\lambda(77 \text{K}) \simeq 0.39 \, \mu$m was found, which is close to $\lambda = 0.42 \, \mu$m used here.
\par

\begin{table} [! h]
\begin{center}
\caption{Best fit parameters}
\label{tab:table1}
\begin{tabular}{|c || c | c | c |  } 
 \hline
 \#   & $I_c$ [$\mu$A] & $\lambda$ [$\mu$m]  & $\sigma$ [$\mu$m]  
\\ 
\hline\hline
S1 & 19.3 & 0.42 & 0.5 \\
\hline
S2 & 20.2 & 0.42 &  0.5  \\ 
\hline
\end{tabular}
\end{center}
\end{table}

While doing the calculations we found that the inductance matrix elements between rows are quite small because the superconducting busbars sufficiently separate the holes of adjacent 1D parallel SQIF arrays (rows). We noticed that neglecting the small row-row mutual inductances made no noticeable difference to any of our results. Therefore, our 2D arrays behave like $N_s$ independent 1D parallel SQIF arrays where the 2D array DC voltage is simply the sum over all the row DC voltages. Being allowed to neglect the row-row coupling simplifies Eq. (16) and dramatically reduced the computational time needed to calculate the time evolution of the JJ phases.
\\
\par
Though each of our 2D SQIF arrays contains $N_s N_p = 1002$ JJs, $N_s$ is still not large enough for $V(B_a)$ and $V(I_b^{tot})$ to become independent of the chosen $I_c$-disorder set \{$\hat{\eta}_{ kk}$\}. We have investigated 40 different randomly chosen sets \{$\hat{\eta}_{kk}$\} for both S1 and S2, and selected the two $I_c$-disorder sets that fitted our S1 and S2 experimental data best.

Figure 5 shows calculated $V(B_a)$ characteristics for S1 and S2, using the parameters in Table III, for 10 different randomly chosen $I_c$-disorder sets \{$\hat{\eta}_{kk}$\} for each array. The $V(B_a)$ curves vary, as each $I_c$-disorder set \{$\hat{\eta}_{kk}$\} produces a somewhat different result. 
Figure 6 displays the $I_c$-disorder set \{$\hat{\eta}_{kk}$\} used to calculate the dotted $V(B_a)$ characteristics of S1 in Fig. 5. Due to the long tail of a log-normal distribution with $\sigma = 0.5$, values of $\hat{\eta}_{kk}$ as large as 4 can be seen. 

\begin{figure}[H]
\begin{center}
\hspace*{-7mm}
\includegraphics[width=0.57\textwidth]{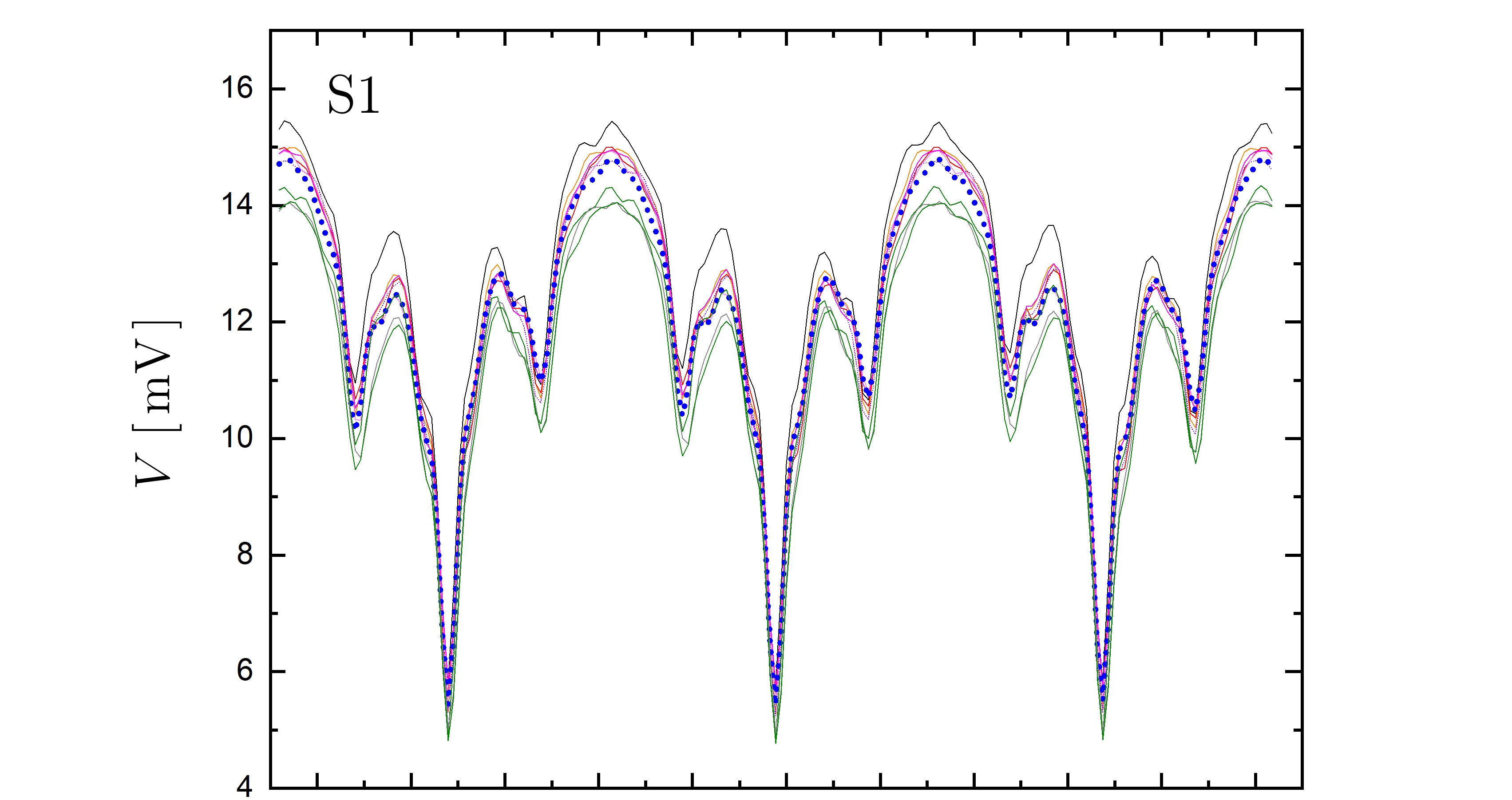}
\label{default}
\hspace*{-7mm}
\includegraphics[width=0.57\textwidth]{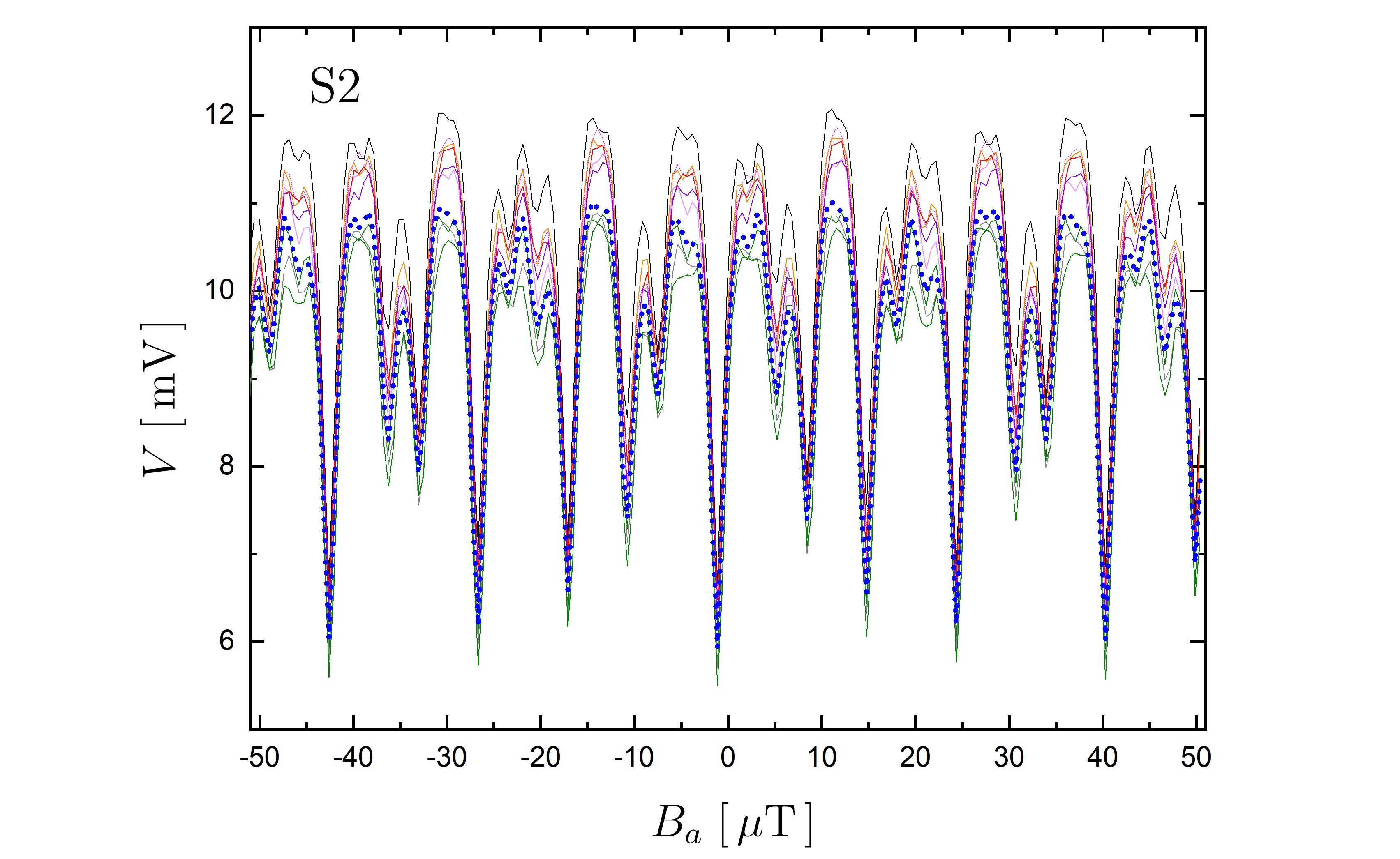}
\caption{$V(B_a)$ characteristics of arrays S1 and S2 for 10 different $I_c$-disorder sets \{$\hat{\eta}_{kk}$\}.}
\label{default}
\end{center}
\end{figure}

\begin{figure}[H]
\begin{center}
\hspace*{-1mm}
\includegraphics[width=0.445\textwidth]{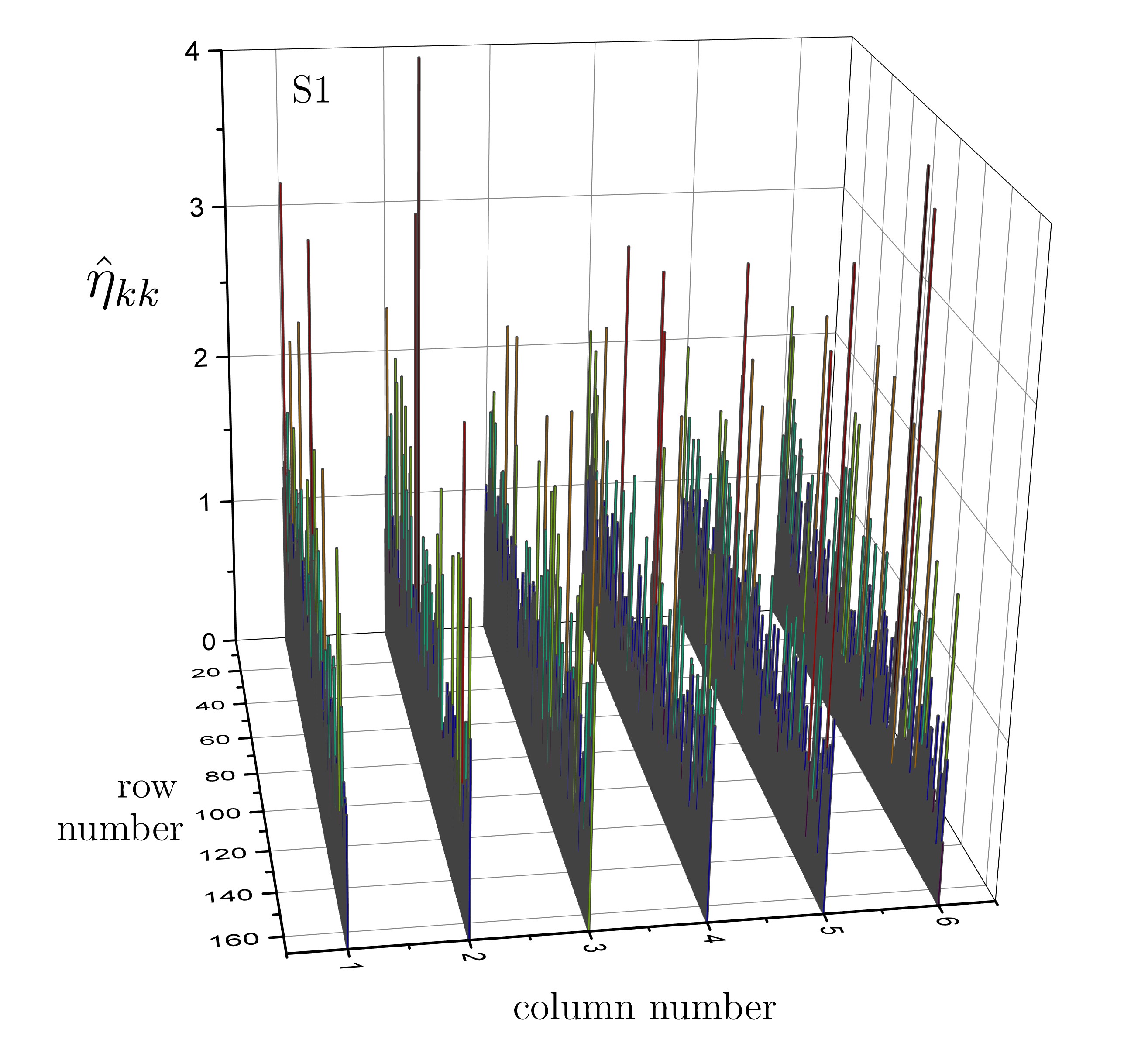}
\caption{$I_c$-disorder set \{$\hat{\eta}_{kk} = I_{ck} / I_c$\} in array S1 used for calculating the dotted curve in Fig. (5). The array has $N_p = 6$ JJ columns and $N_s = 167$ JJ rows.}
\label{default}
\end{center}
\end{figure}

The calculated dotted curves from Fig. 5 for S1 and S2, together with our experimental S1 and S2 data from Fig. 2, are displayed in Fig. 7. Figure 7 shows a key result of our paper. The quantitative agreements between experimental data and calculation for both S1 and S2 are exceptional. Still, we could have improved the agreement even further by searching for more optimal $I_c$-disorder sets \{$\hat{\eta}_{kk}$\} . It turns out that the effective hole areas $A_m^{\text{eff}}$ are important ingredients to describe the experimental $V(B_a)$ data correctly. As can be seen in Fig. 7, the calculated curves are slightly more stretched along the $B_a$ axis than the experimental data. After the calculation, we found out that this discrepancy can be partially explained by the fact that the layout size of the arrays was actually about 2\% larger than the values given in Table I. Furthermore, our calculations showed that the term in Eq. (16) which contains the bias current inductance $\vec{L}^{I_b}$ can be neglected without affecting our calculated $V(B_a)$ characteristics.

\begin{figure}[! h]
\begin{center}
\hspace*{-7mm}
\includegraphics[width=0.60\textwidth]{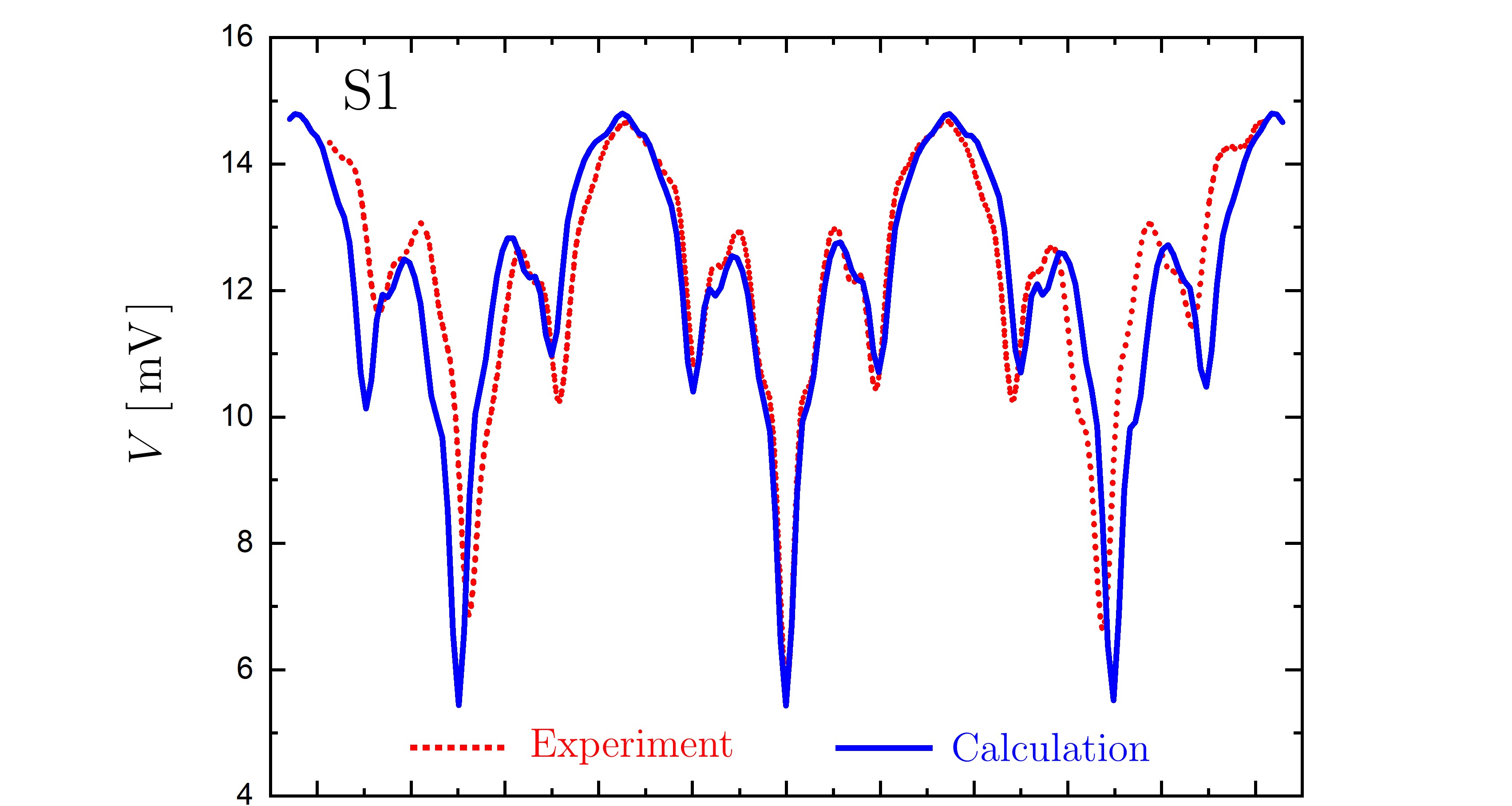}
\label{default}
\hspace*{-7mm}
\includegraphics[width=0.60\textwidth]{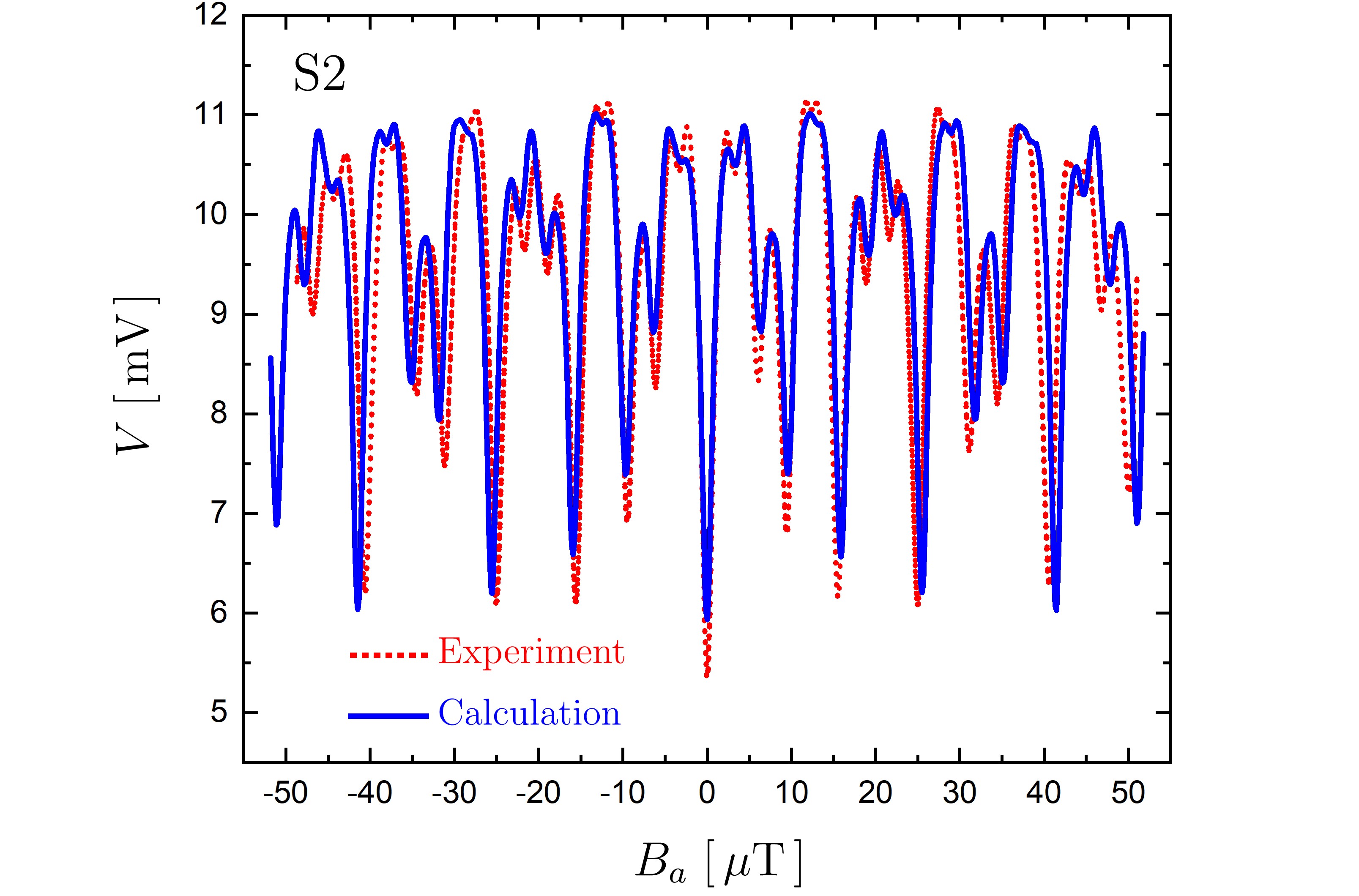}
\caption{Comparison between experimental and calculated $V(B_a)$ characteristics for array S1 and S2.}
\label{default}
\end{center}
\end{figure}

To gain a better understanding of the role of the London penetration depth $\lambda$, Fig. 8 shows, for S1 and S2, how $V(B_a)$ (for positive $B_a$) changes with $\lambda$ while all other parameters stay unchanged. Increasing $\lambda$ causes the voltage modulation depth $\Delta V = V_{max} -V_{min}$ to decrease. This is mainly due to the increase in the kinetic inductances which are proportional to $\lambda^2$. This is similar to a DC-SQUID where an increase of the screening parameter $\beta_L$ leads to a decrease of $\Delta V$ \cite{TES77}.

\begin{figure}[h]
\begin{center}
\hspace*{-7mm}
\includegraphics[width=0.60\textwidth]{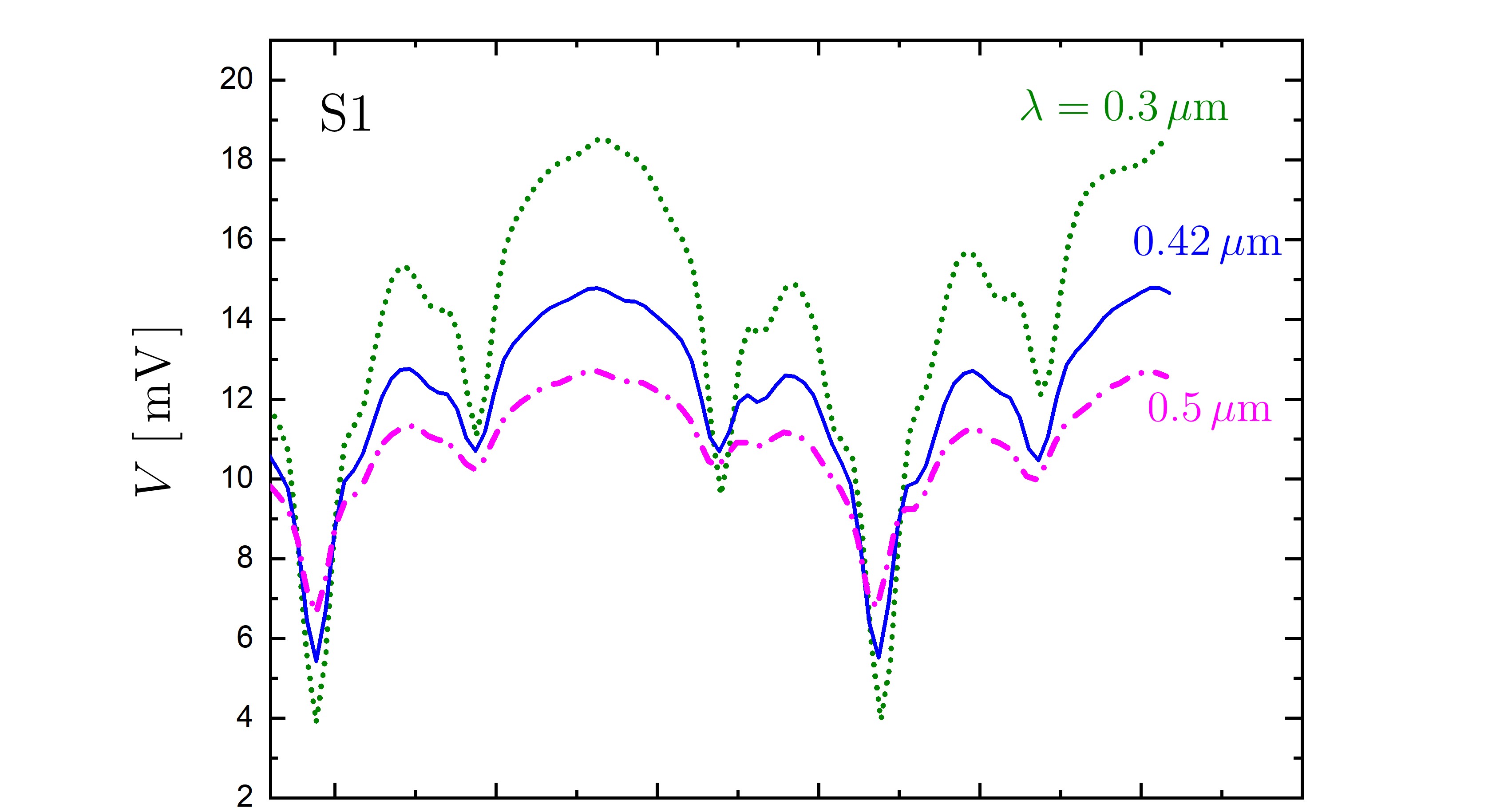}
\label{default}
\hspace*{-7mm}
\includegraphics[width=0.60\textwidth]{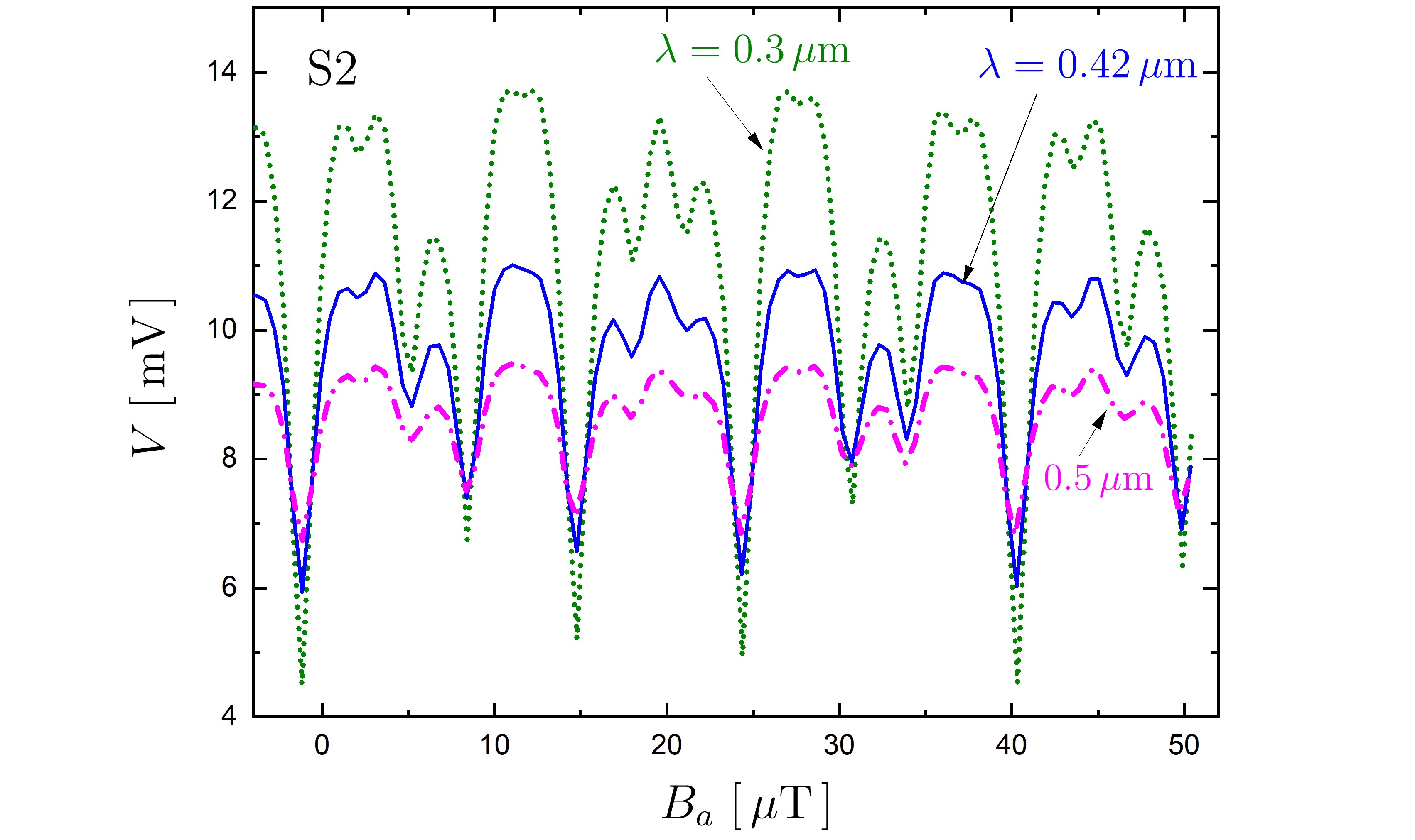}
\caption{Calculated $V(B_a)$ characteristics of array S1 and S2 for three different values of $\lambda$.}
\label{default}
\end{center}
\end{figure}

Figure 9 shows how rapidly the voltage modulation depth $\Delta V(\lambda)$ for S1 and S2, which is a measure of the maximum transfer function $dV / dB_a |_{max}$ \cite{GAL22}, decreases with increasing $\lambda$. Note that $\lambda$ enters our calculations via the factor $\lambda^2 / d$ (Eq. (24), and Eq. (30) with Eq. (5)), which is known as the Pearl penetration depth \cite{PEA64}. Thus, increasing the film thickness $d$, as long as $\lambda > d$ \cite{CLE05,BRA05}, would further increase the modulation depth $\Delta V$ of our arrays, assuming that all other parameters stay unchanged.
Figure 10 shows, for S1 and S2, how the $V(B_a)$ characteristics (for positive $B_a$) change with the $I_c$-spread $\sigma$ while all other parameters stay unchanged. Increasing $\sigma$ smoothes the $V(B_a)$ curves, which decreases the voltage modulation depth $\Delta V$. The underlying cause

\begin{figure}[H]
\begin{center}
\hspace*{0mm}
\includegraphics[width=0.50\textwidth]{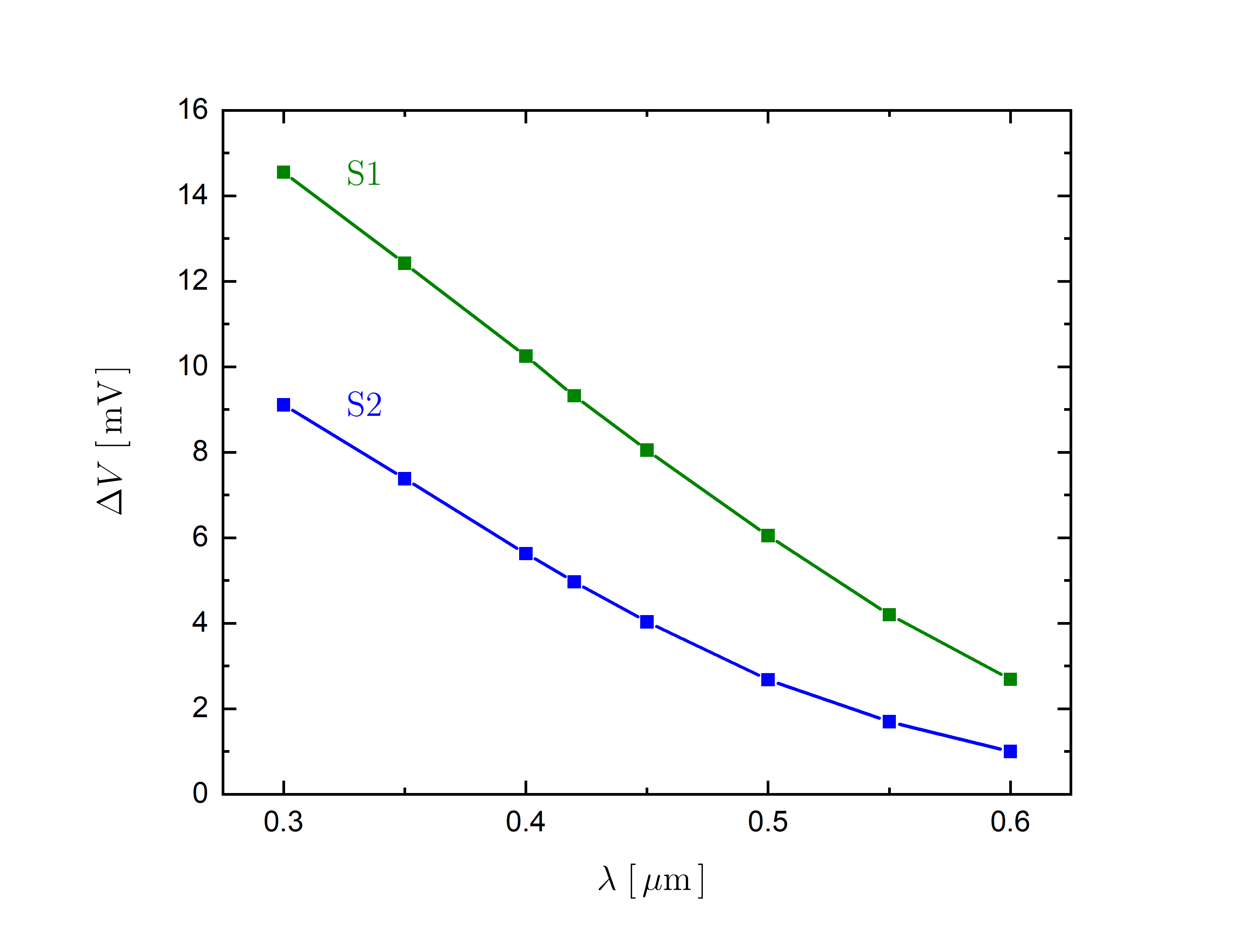}
\caption{Calculated voltage modulation depth $\Delta V$ versus $\lambda$ of array S1 and S2.}
\label{default}
\end{center}
\end{figure}

\begin{figure}[H] 
\begin{center}
\hspace*{-7mm}
\includegraphics[width=0.60\textwidth]{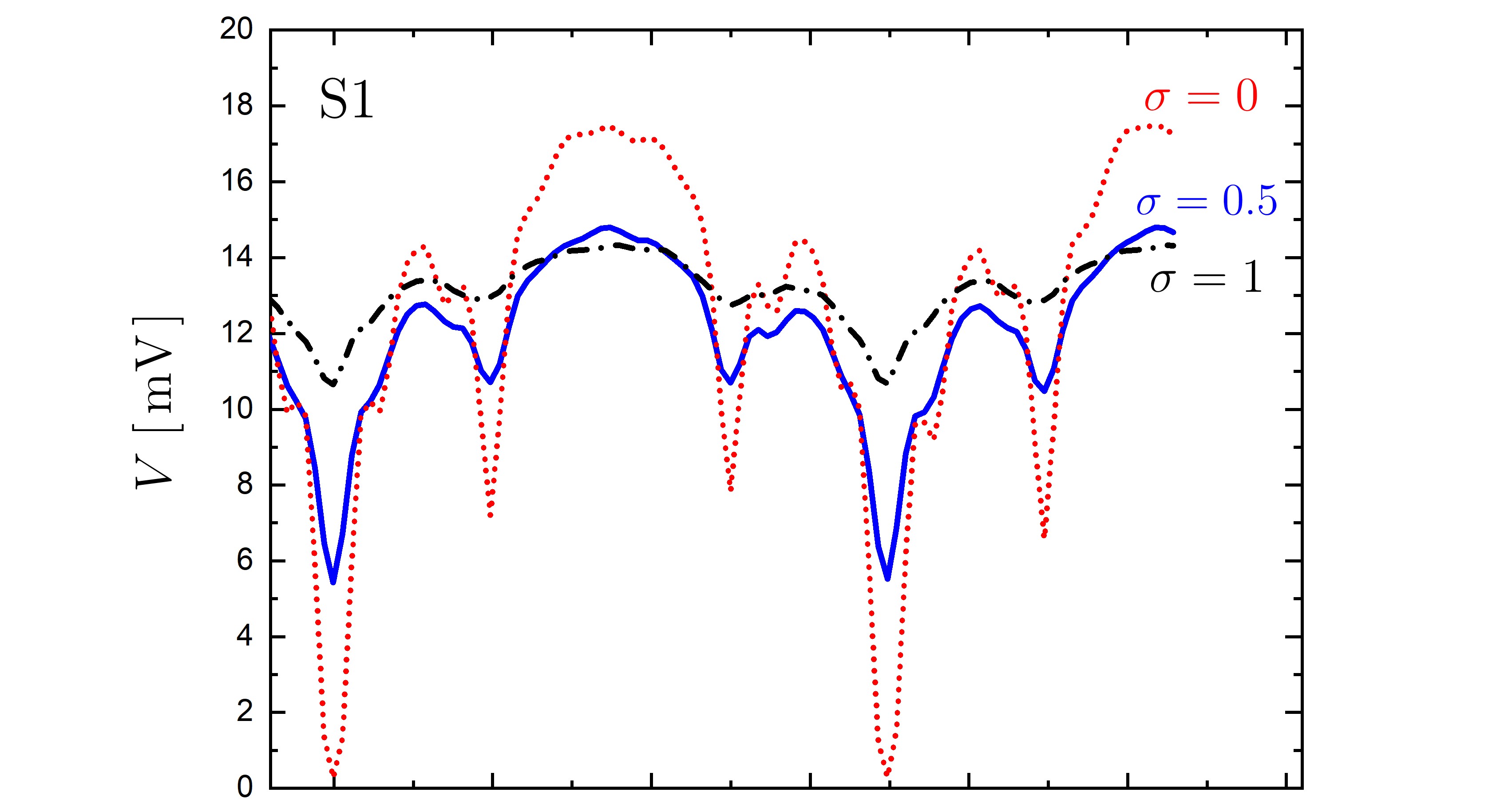}
\label{default}
\hspace*{-7mm}
\includegraphics[width=0.60\textwidth]{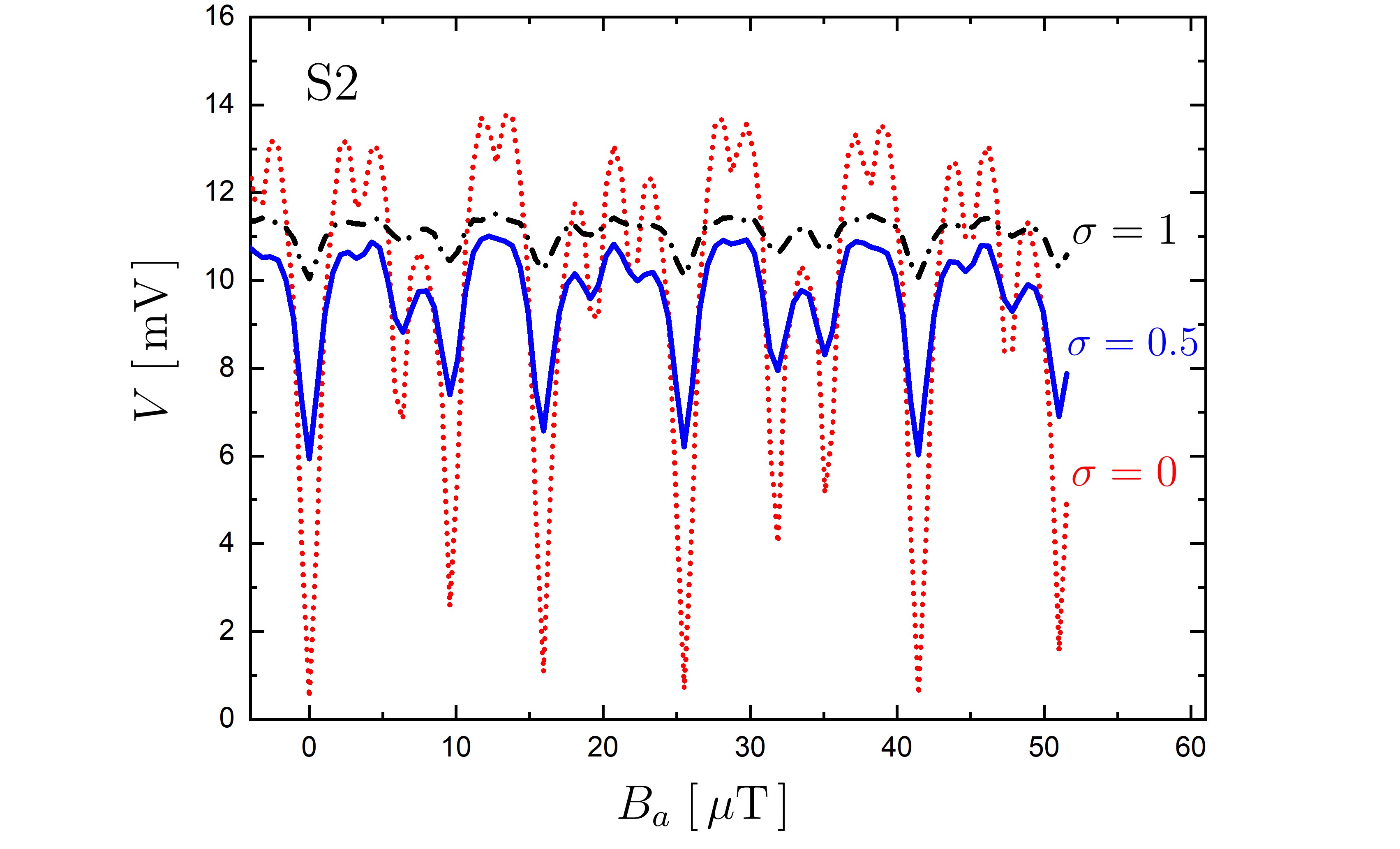}
\caption{Calculated $V(B_a)$ characteristics of array S1 and S2 for three different $I_c$-spreads $\sigma$.}
\label{default}
\end{center}
\end{figure}

\noindent
is revealed by looking at the $V(\Phi^{\text{(app)}})$ characteristics of the simple DC-SQUID ($N_s = 1, N_p = 2$) with $R_1 \neq R_2$ and $I_{c1} \neq I_{c2}$ \cite{TES77, MUL01}. Here, the $R$-asymmetry causes a skewing of the $V(\Phi^{\text{(app)}})$ response \cite{TES77} and the $I_c$-asymmetry a $\Delta\Phi^{\text{app}}$-shifting to the left or right by an amount $\Delta \Phi^{\text{(app)}}  = L_s \, | \, I_{c1} - I_{c2} \, | \, / \, 2$ where $L_s$ is the self-inductance of the DC-SQUID loop \cite{MUL01}. In addition to the reflection symmetry breaking, the modulation depth decreases with increasing $I_c$-asymmetry \cite{TES77, MUL01}. The modulation depth is less affected by the $R$-asymmetry. In our 2D SQIF arrays each row has 6 JJ's in parallel and 5 holes. As the $I_{ck}$'s and $R_k$'s are all different, this produces in each row a different reflection asymmetry, {\it{i.e.}} $V_{row}(B_a) \neq V_{row}(-B_a)$, due to different skewing and $B_a$-shifting. Because of the very weak inductive coupling between rows, as mentioned above, the time averaged voltages $V_{row}(B_a)$ across rows can simply be added up to give the total array voltage $V(B_a)$. Adding up a large number of different row voltages $V_{row}(B_a)$ causes a $V(B_a)$ smoothing which reduces the modulation depth $\Delta V$ of the array but also reduces the reflection asymmetry.

Figure 11 displays $\Delta V(\sigma)$ normalized by $\Delta V(\sigma = 0)$. The figure shows that an $I_c$-spread $\sigma = 0.5$ reduces $\Delta V(\sigma) / \Delta V(0)$ in the case of S1 from 1 to about 0.55 and in the case of S2 from 1 to about 0.4.
Figure 11 reveals that $I_c$-$R$-disorder has a stronger detrimental effect on $\Delta V$ in an array with larger inductances like S2 than in an array with smaller inductances like S1. Figure 11 is one of the key results of this paper.

\begin{figure}[! h]
\begin{center}
\hspace*{-4mm}
\includegraphics[width=0.55\textwidth]{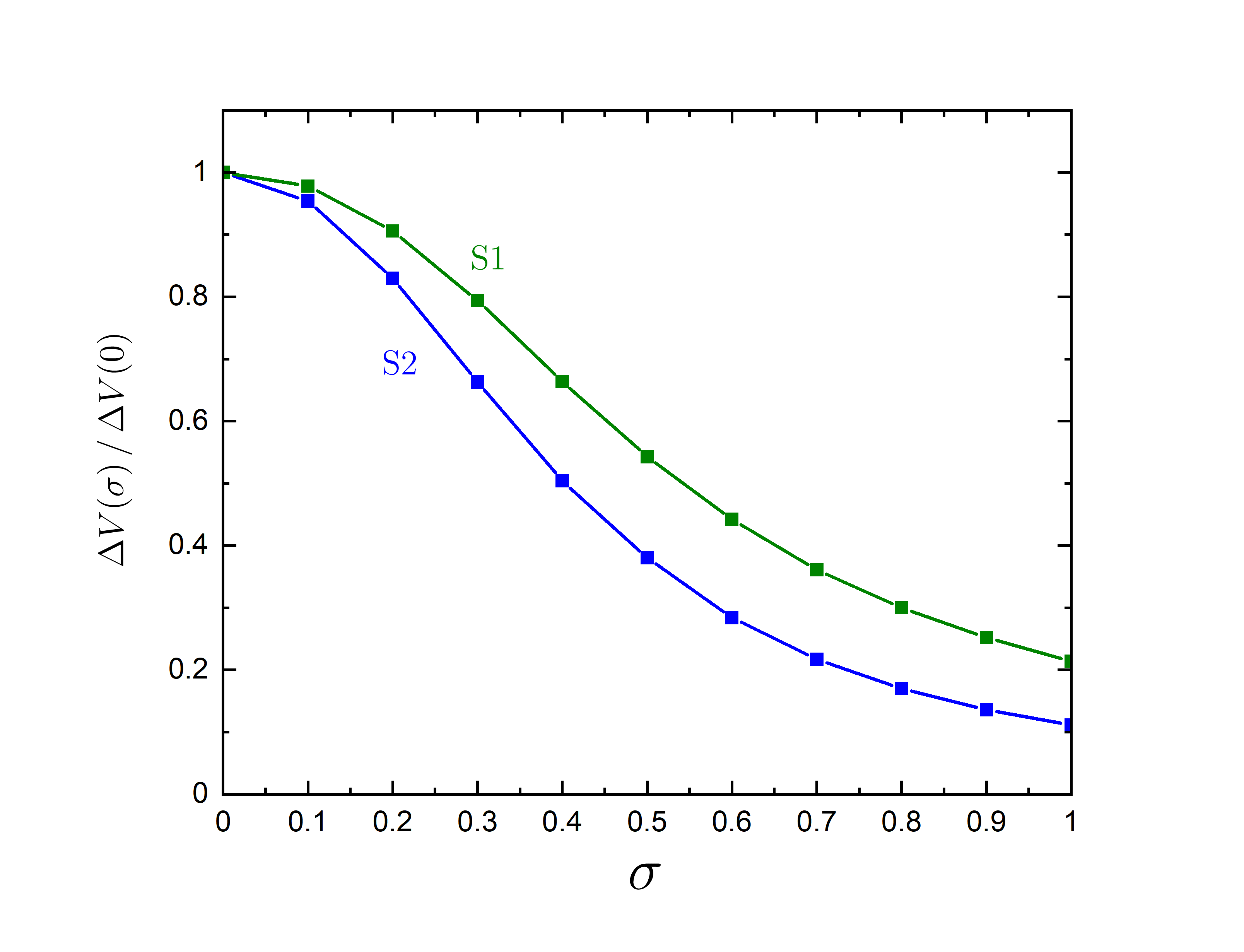}
\caption{The calculated normalized voltage modulation depth $\Delta V(\sigma) \, / \, \Delta V(0)$ versus $I_c$-spread $\sigma$ for array S1 and S2.}
\label{default}
\end{center}
\end{figure}

In Appendix D we discuss the weak reflection asymmetry of $V(B_a)$.
\\
\par

Figure 12 shows the experimental and calculated $V(I_b^{tot})$ characteristics in zero applied magnetic field for S1 an S2. The calculation does not perfectly reproduce the experimentally observed $V(I_b^{tot})$ at low $I_b^{tot}$. Better agreement can possibly be achieved by modifying the log-normal distribution $p(\eta)$ in Eq. (31) such that low $I_c$ values become more emphasized. The values of $N_p I_c$ for S1 and S2 are indicated in Fig 12.  In contrast to $V(B_a)$, the calculated shape of $V(I_b^{tot})$ is not very sensitive to a particular choice of the $I_c$-disorder set \{$\hat{\eta}_{kk}$\} for a given $\sigma$, but is very sensitive to the choice of $I_c$, {\it{i.e.}} the average $I_{ck}$ value.

Figure 13 again displays $V(I_b^{tot})$ for S1 but here the calculated additional dotted curves reveal how the parameters $T$ and $\sigma$ affect the shape of $V(I_b^{tot})$. Going from curve (1) to (2) shows the effect of temperature $T = 77$ K, and going from (2) to (3) the effect of $\sigma = 0.5$.

\begin{figure}[H]
\begin{center}
\hspace*{-7mm}
\includegraphics[width=0.55\textwidth]{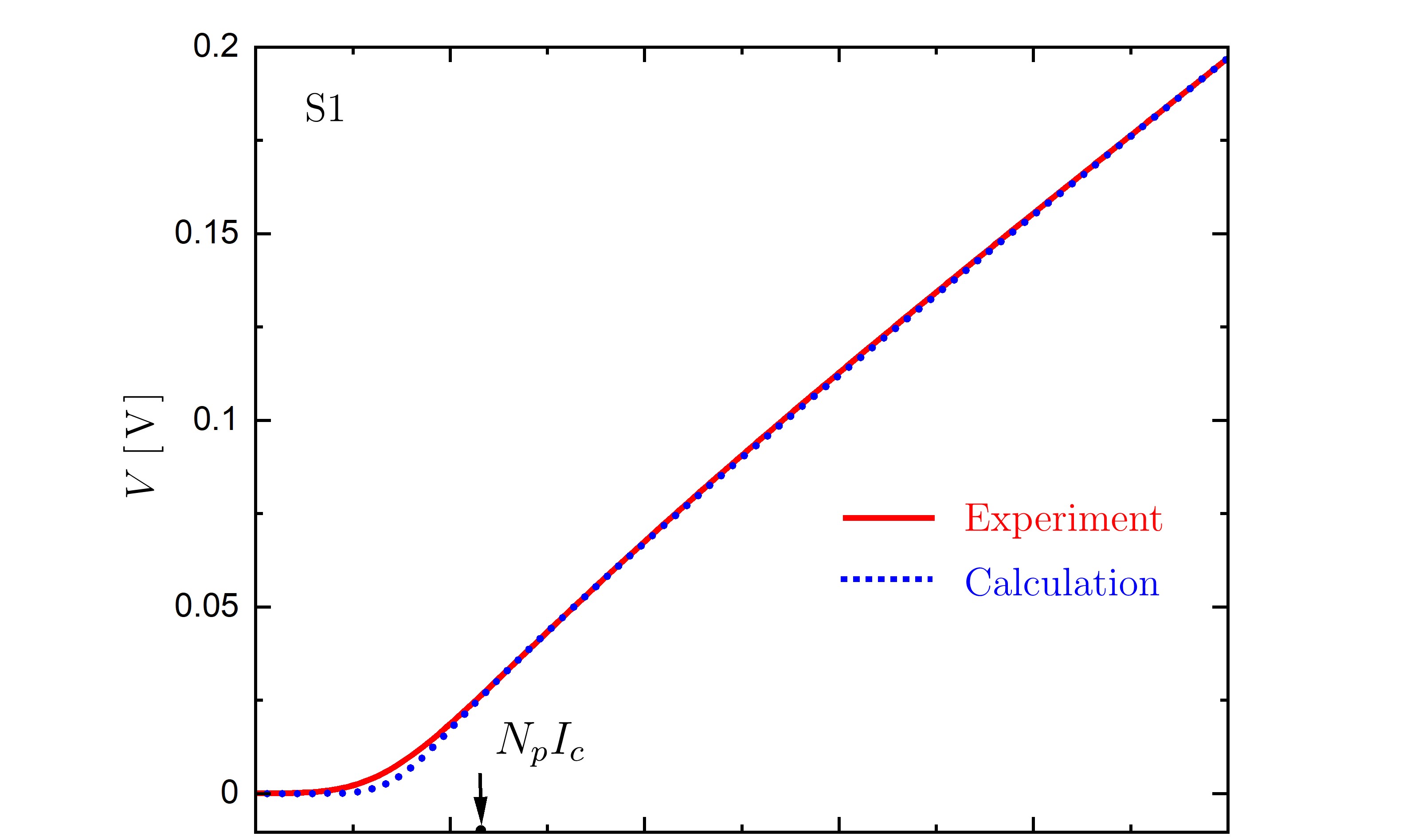}
\label{default}
\hspace*{-7mm}
\includegraphics[width=0.55\textwidth]{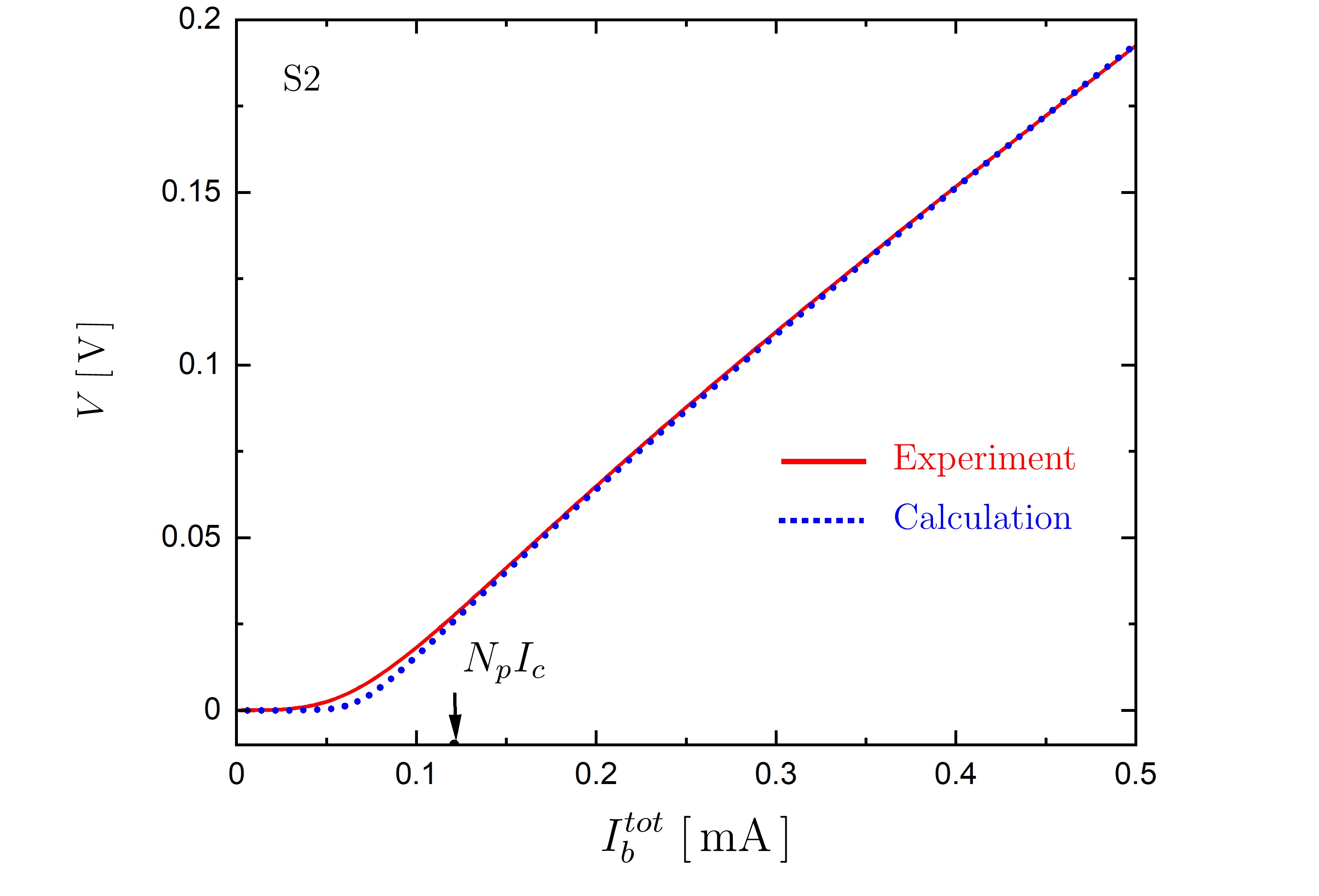}
\caption{Experimental and calculated $V(I_b^{tot})$  characteristics of array S1 and S2.}
\label{default}
\end{center}
\end{figure}

\begin{figure}[H]
\begin{center}
\hspace*{0mm}
\includegraphics[width=0.46\textwidth]{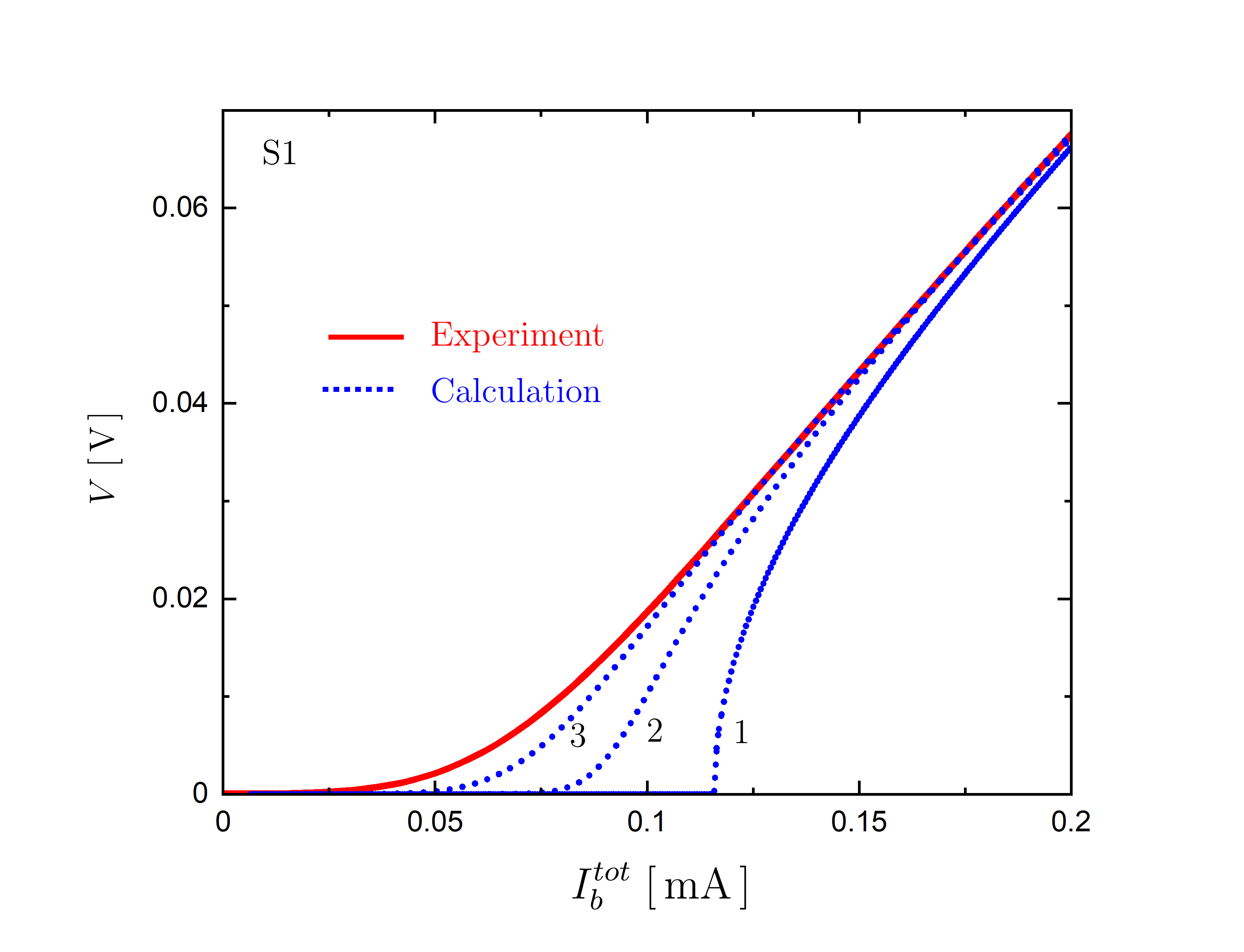}
\caption{Experimental and calculated $V(I_b^{tot})$ characteristics of array S1. The parameters used to calculate the 3 dotted blue curves are (1) $T = \sigma =  0$, (2) $T = 77$ K, $\sigma = 0$, and (3) $T = 77$ K, $\sigma = 0.5$.}
\label{default}
\end{center}
\end{figure}

\section{Summary and Conclusion}

We measured the $V(B_a)$ characteristics at optimal bias currents and the $V(I_b^{tot})$ characteristics at $B_a = 0$ of two 2D SQIF arrays with $N_s = 167$ and $N_p = 6$. 

To model our experimental data, we developed a theoretical model that takes into account the Johnson noise in the JJ resistances $R_k$, the spreads in $I_{ck}$ and $R_k$, and the wide, thin-film busbar structure of the arrays. The model was made numerically efficient by dividing the superconducting current density into its Meissner current, circulating current, and bias current parts. Using the stream function approach for $\lambda > d$, we calculated the effective areas of the array holes, inductance matrices and bias current inductance vectors. The calculations revealed that the busbar height was sufficiently large such that the inductive row-row coupling could be neglected. This significantly simplified the numerical calculations and made the investigation of the $I_c$-$R$-disorder of our $N_s = 167$ arrays computationally possible. 

Empirical findings that $R_k$ can be expressed in terms of $I_{ck}$ reduced the parameter space, and we could identify $\{I_{ck}\}$ sets from a log-normal distribution that produced $V(B_a)$ characteristics that agreed well with our experimental data.

Our calculations demonstrated that the voltage modulation depth $\Delta V$, which is a measure for the maximum transfer function, decreases with increasing $\lambda$ due to the increase in kinetic inductances which are proportional to $\lambda^2 / d$.

Most of all, our calculations showed how $\Delta V$ decreases with increasing $I_c$-spread $\sigma$, where the decrease was more rapid for the high inductance array S2. Our experimental data were best described by $\sigma = 0.5$, showing that the $I_c$-$R$-disorder in our arrays reduced $\Delta V$ by 45\% in S1 and 60\% in S2.

We observed slight reflection asymmetries in our experimental $V(B_a)$ characteristics, caused by the $I_c$-$R$-disorder, and our model predicted a similar degree of asymmetry as observed experimentally. Generally, the refection asymmetry increases with increasing $I_c$-spread $\sigma$, but decreases when $N_s$ is increased. 

The calculated $V(I_b^{tot})$ characteristics did not fully agree with our experimental data at low $I_b^{tot}$, possibly due to the choice of a log-normal distribution that did not sufficiently emphasize low $I_{ck}$ values. Modifying our log-normal distribution would give better agreement with $V(I_b^{tot})$ without affecting much the $V(B_a)$ results. Our calculations also showed how the $V(I_b^{tot})$ characteristic changes with temperature ($T = 0$ to $T = 77$ K) and with $\sigma$ ($\sigma = 0$ to $\sigma = 0.5$).  

Our findings suggest that the use of YBCO thin films in SQIF arrays is promising, as the critical current disorder inherent in their fabrication process reduces the transfer function in our case only by about a factor of 2. This is one of our important results. Our study highlights the importance of theoretical modeling in understanding the performance of superconducting devices and provides insights that could inform the design of future SQIF arrays. While our detailed theoretical modeling provided valuable insight into the role of $I_c$-$R$-disorder for the specific arrays tested, more general investigations are needed to fully understand its role as a function of screening parameter $\beta_L$, noise strength $\Gamma$, kinetic inductance fraction, $N_s$ and $N_p$ values, and different array planar geometries.
\\
\par

\begin{acknowledgments}
The authors would like to thank Marc Gali Labarias for stimulating discussions.
\end{acknowledgments}

\appendix

\newpage

\section{Effective hole areas}

By choosing any constant applied magnetic field, $B_a > 0$, and the boundary condition $g |_{\partial \Omega} = 0$, for all boundaries $\partial \Omega$, one can calculate numerically, by solving Eq. (24), the part of $g(x,y)$ that corresponds to the Meissner shielding current density $j^{(Mei)}$. In this case $f_s = 0$ (Eq. (25)). Stream-lines of $j^{(Mei)}$ in array S1 and S2 are shown in Fig. 14. Using Eqs. (28) and (30), the fluxoids $\Phi_{1,m}$ (Eq. (9)) of the different holes are determined and the effective hole area enhancement factors $A_m^{\text{eff}} / A_m$ (Eq. (13)) are calculated.

\begin{figure}[H]
\begin{center}
\hspace*{-8mm}
\includegraphics[width=0.375\textwidth]{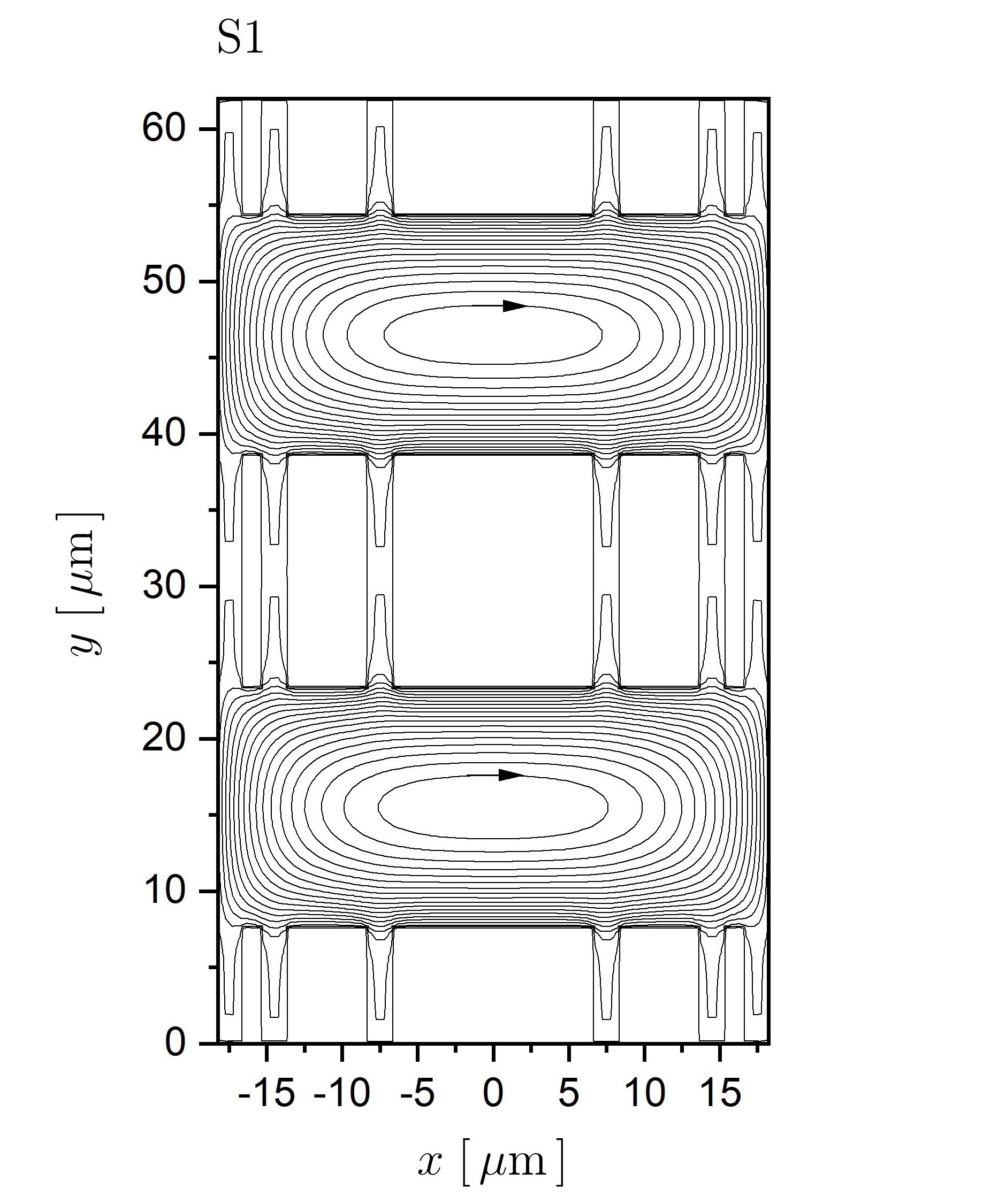}
\label{default}
\hspace*{-8mm}
\includegraphics[width=0.55\textwidth]{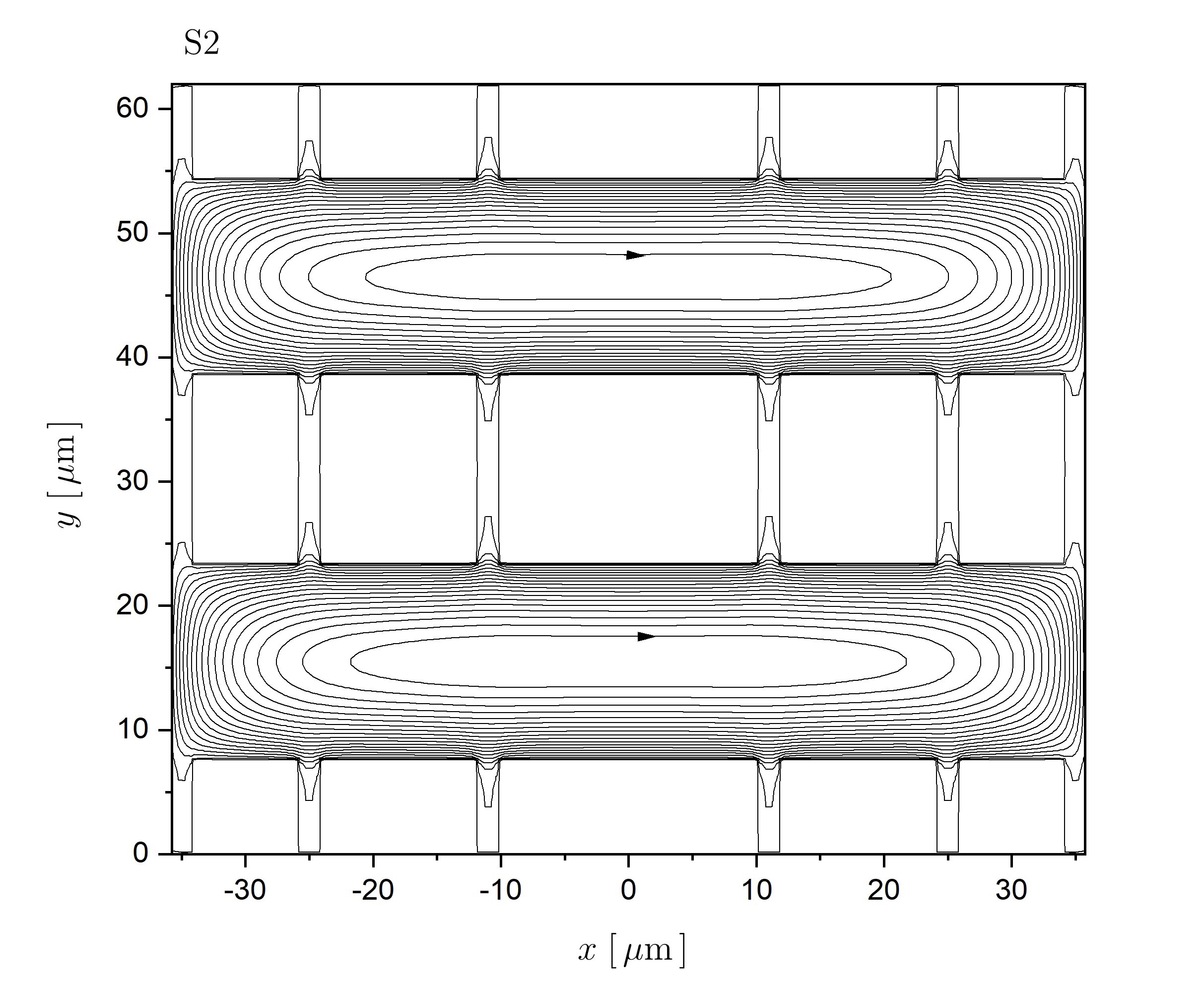}
\caption{Meissner current streamlines in S1 and S2.}
\label{default}
\end{center}
\end{figure}

\noindent
$A_m^{\text{eff}} / A_m$ for different holes $m$ are displayed in Fig. 15, where the geometric and kinetic parts are indicated. Figure 15 shows that the kinetic part plays an important role.

\begin{figure}[H]
\begin{center}
\hspace*{-6mm}
\includegraphics[width=0.50\textwidth]{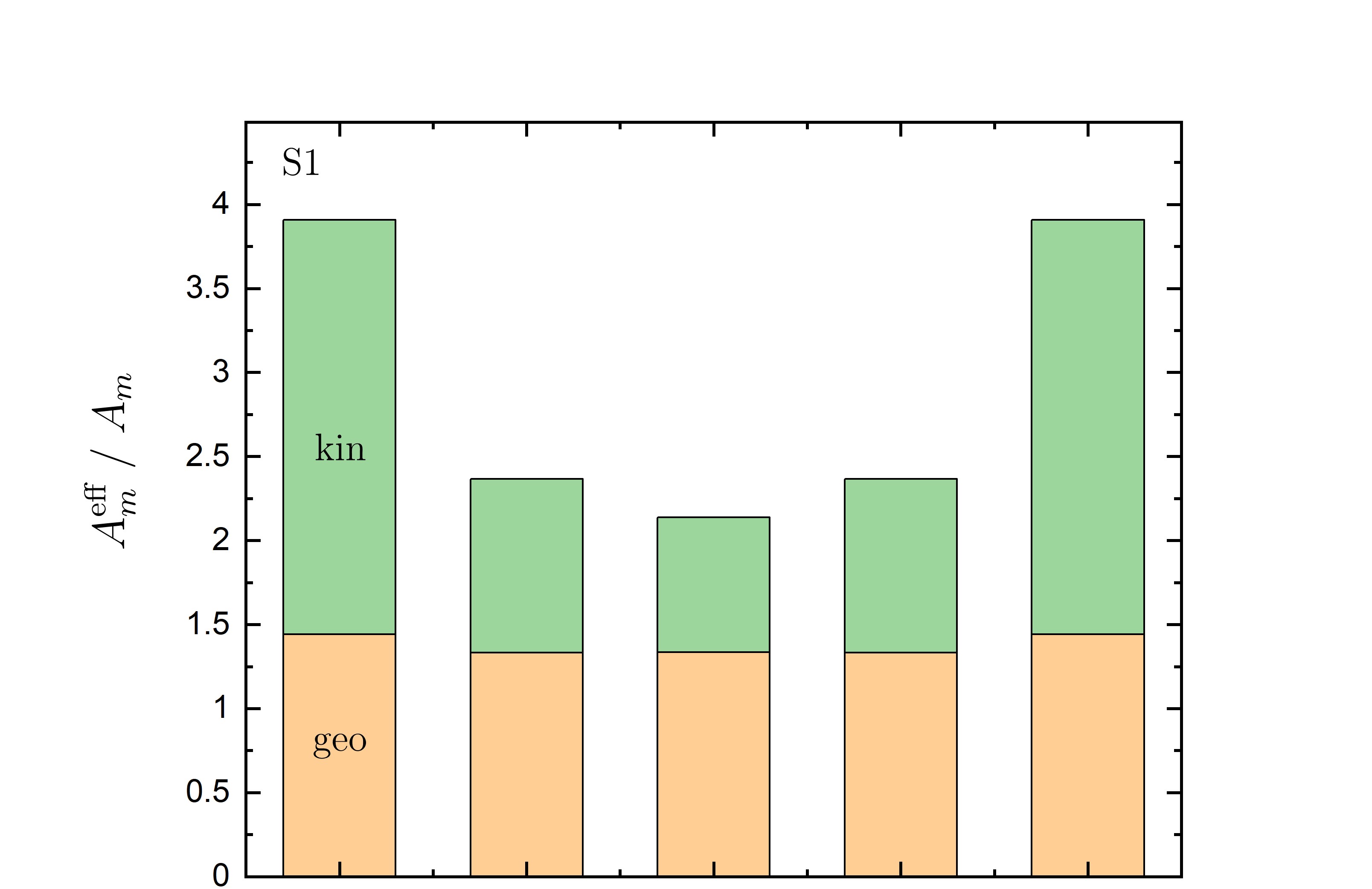}
\label{default}
\hspace*{-6mm}
\includegraphics[width=0.50\textwidth]{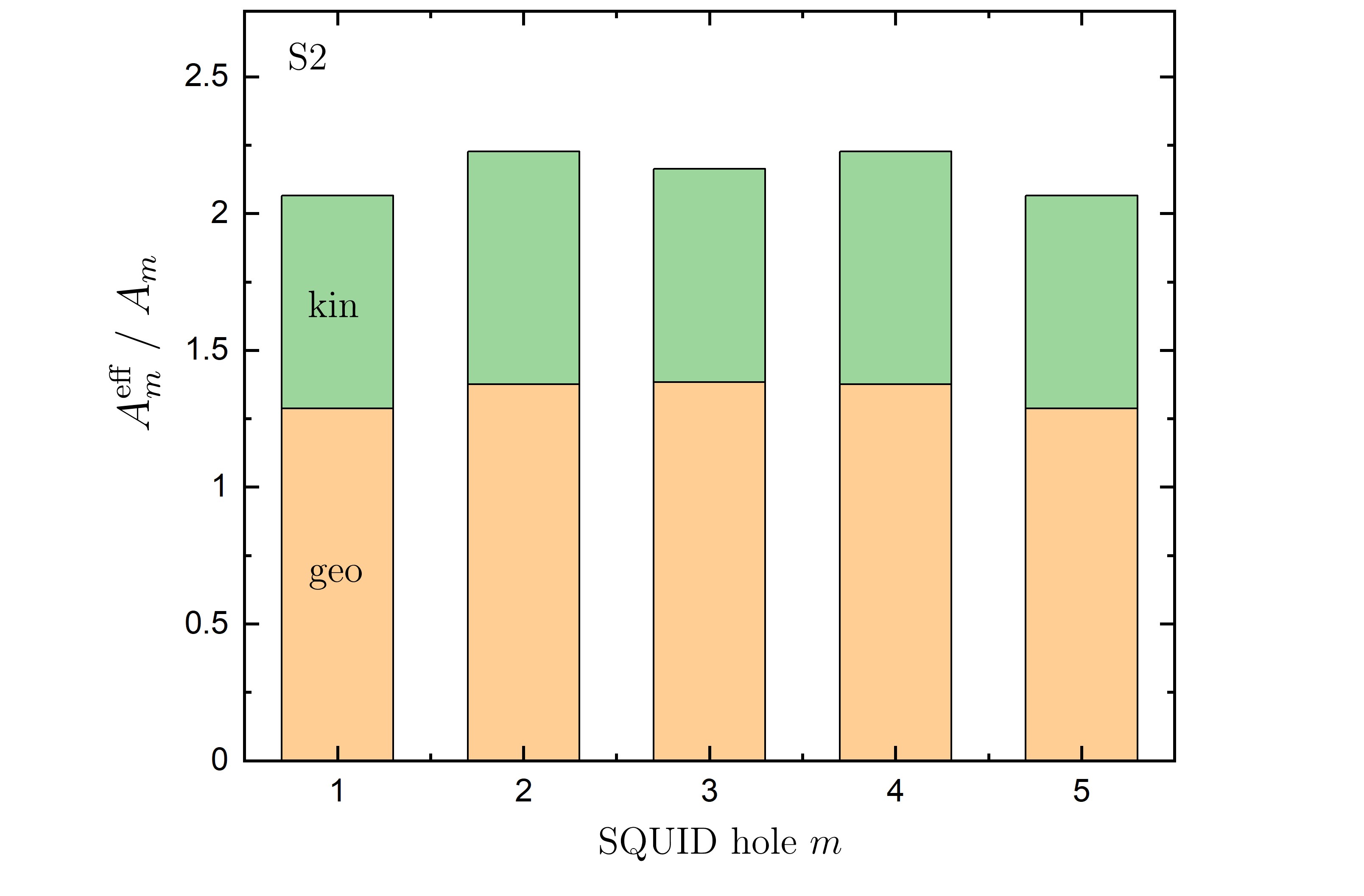}
\caption{Effective hole area enhancement factor $A_m^{eff} / A_m$ of S1 and S2. Geometric and kinetic parts are indicated.}
\label{default}
\end{center}
\end{figure}

\section{Inductance matrix $\hat{L}$}

One can determine the part of $g(x,y)$, which corresponds to a counterclockwise circulating current $J_m$ around hole $m$, by choosing $B_a = 0$ and solving Eq. (24) numerically with the appropriate boundary condition. The boundary condition is $g|_{\Omega_m} = J_m$ ($J_m > 0$ but arbitrary), and $g$ increases linearly from 0 to $J_m$ along the left JJ and decreases linearly from $J_m$ to 0 along the right JJ. Everywhere else $g|_{\partial \Omega} = 0$. In this case, $f_s$ in Eq. (25) can be obtained analytically as $f_s(x,y) = [ \, P^u_m(x,y) + P^u_m(x,-y) \, ] \, J_m$, where the formulas for $P^u_m$ can be found in Appendix B of ref. \cite{MUL21}. Examples of current stream-lines of currents flowing around a hole $m = 2$ in S1 and S2 are shown in Fig. 16. One can use Eqs. (28) and (30) to determine the fluxoids $\Phi_{2,m}$ (Eq. (10)) generated in the different holes, and then with Eq. (14) all the matrix elements of the inductance matrix $\hat{L}$ are calculated. The matrix elements $\hat{L}_{ij}$ of S1 and S2, for the holes in a single row, are plotted in Fig. 17 and their geometric and kinetic parts are indicated. The large self-inductances are seen along the diagonal while the negative mutual inductances are off-diagonal. The row-row mutual inductance matrix elements are found to be very small and are not displayed here. Figure 17 shows that the kinetic part plays a dominant role.

\begin{figure}[H]
\begin{center}
\hspace*{-3mm}
\includegraphics[width=0.375\textwidth]{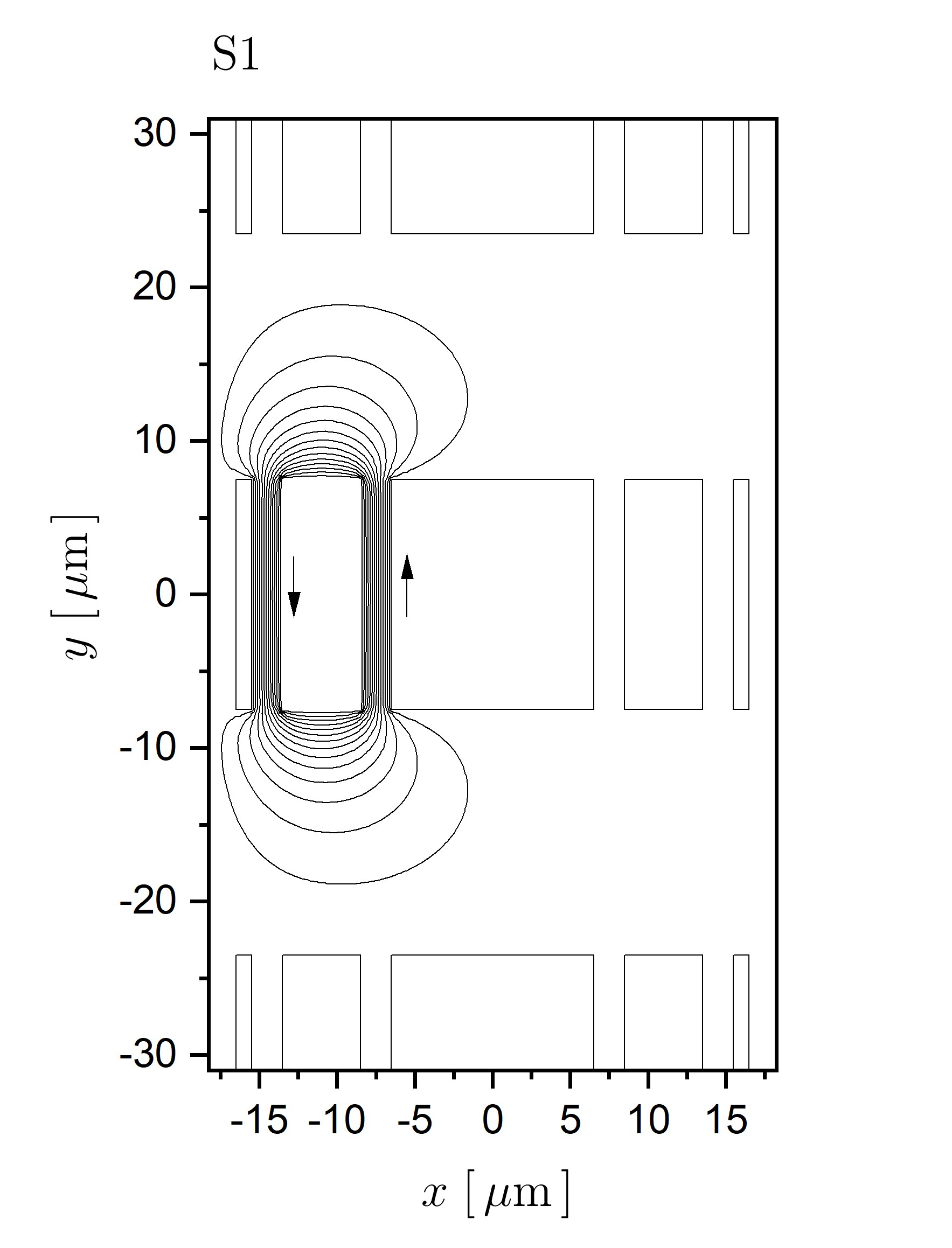}
\label{default}
\hspace*{-3mm}
\includegraphics[width=0.55\textwidth]{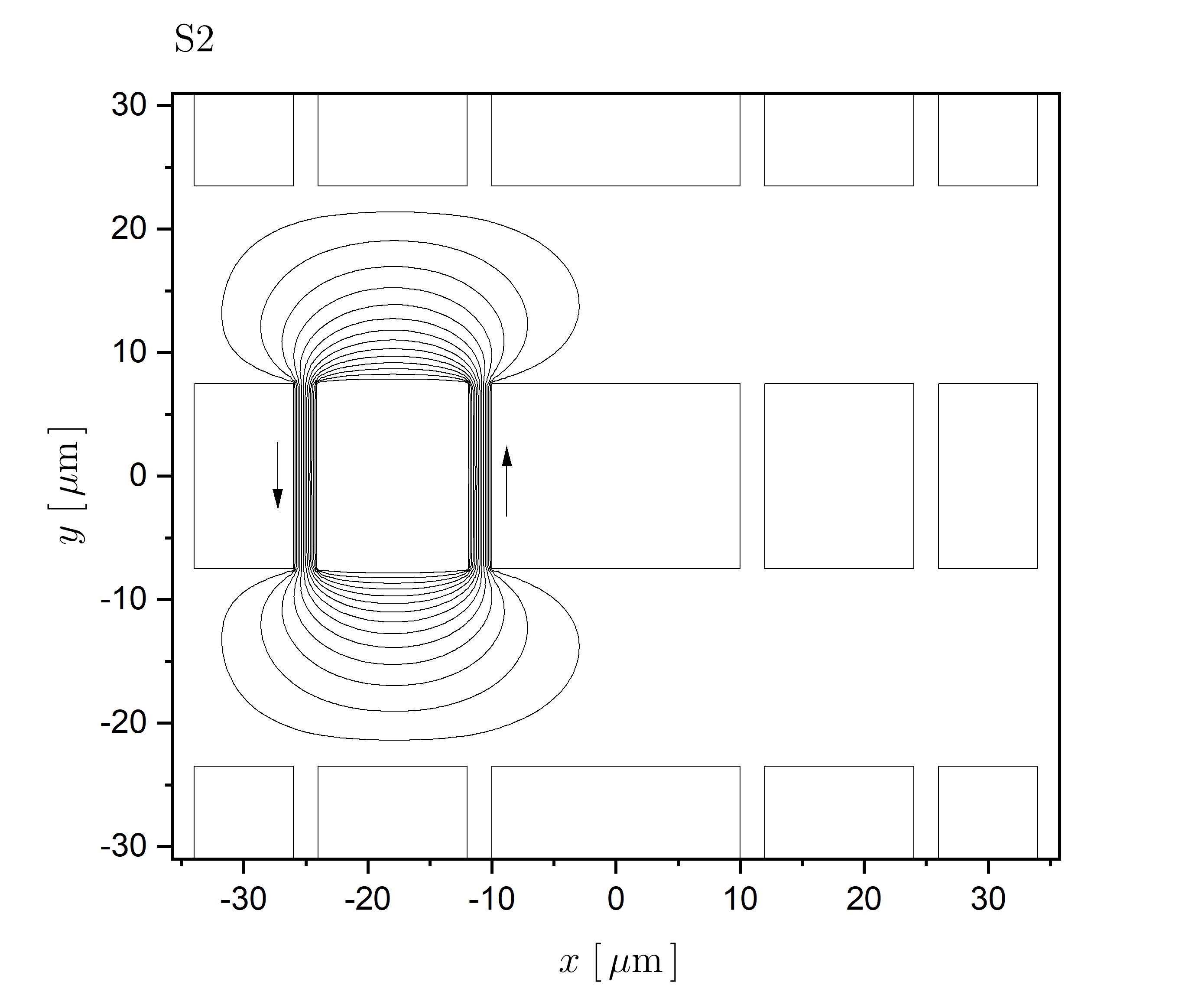}
\caption{Current streamlines in array S1 and S2 to determine the inductance matrix $\hat{L}$.}
\label{default}
\end{center}
\end{figure}

\begin{figure}[! h]
\begin{center}
\hspace*{-7mm}
\includegraphics[width=0.60\textwidth]{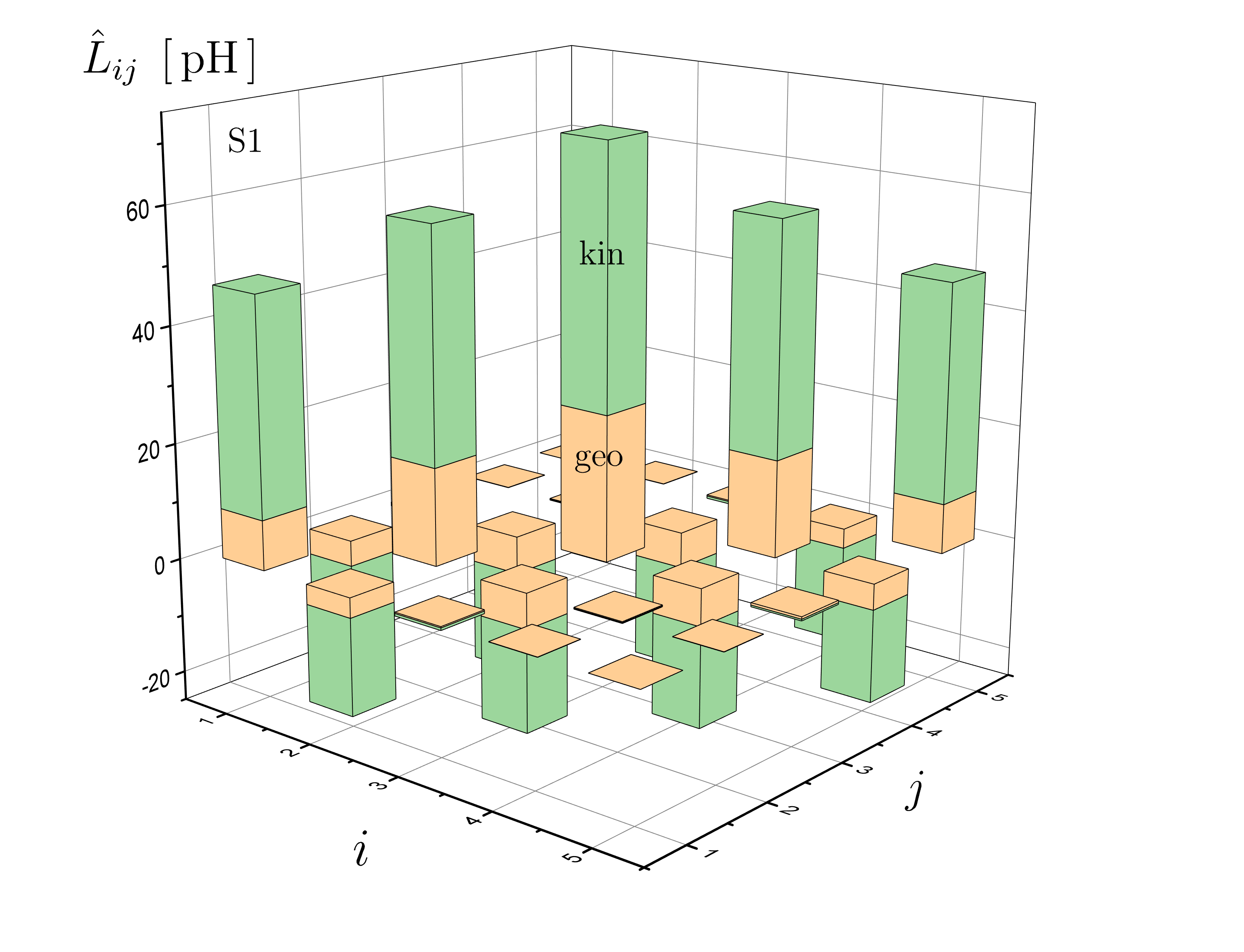}
\label{default}
\hspace*{-7mm}
\includegraphics[width=0.60\textwidth]{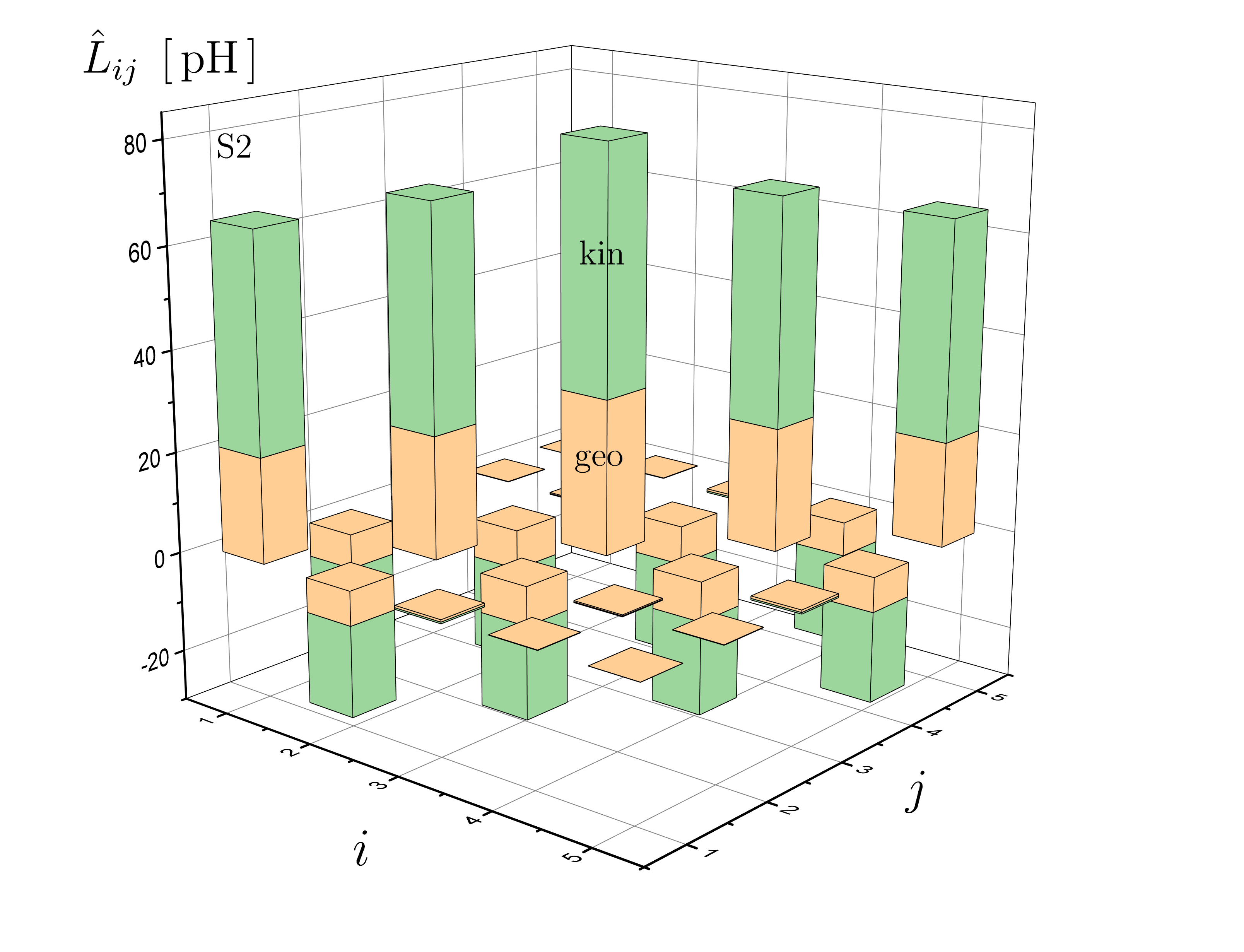}
\caption{Inductance matrix elements $\hat{L}_{ij}$ of a single row, $i = 1$ to 5 and $j = 1$ to 5, for S1 and S2. The geometric and kinetic inductance parts are shown in different colours.}
\label{default}
\end{center}
\end{figure}

\section{Inductance vector $\vec{L}^{I_b}$}

One can determine the part of $g(x,y)$ that corresponds to the bias transport current density $\vec{j}^{(I_b)}$ by choosing $B_a = 0$ and solving Eq. (24) numerically with the appropriate boundary condition. The boundary condition is $g = \pm N_p I_b / 2$ along the left/right outer edges of the array, and, along the inner edges of the holes, $g = - N_p I_b / 2 + i I_b$ with $i = 1$ to $N_p - 1$, while $g$ increases linearly along the JJ's between neighboring holes. Again, $f_s$ can be obtained analytically and can be expressed in terms of $P^u_m(x,y)$ \cite{MUL21}. Streamlines of $\vec{j}^{(I_b)}$ in S1 and S2 are shown in Fig. 18. Using Eqs. (28) and (30), the fluxoids $\Phi_{3,m}$ (Eq. (11)) of the different holes can be determined, and with Eq. (15) the components $L_m^{I_b}$ of the inductance vector $\vec{L}^{I_b}$ are calculated.

\begin{figure}[! h]
\begin{center}
\hspace*{-8mm}
\includegraphics[width=0.375\textwidth]{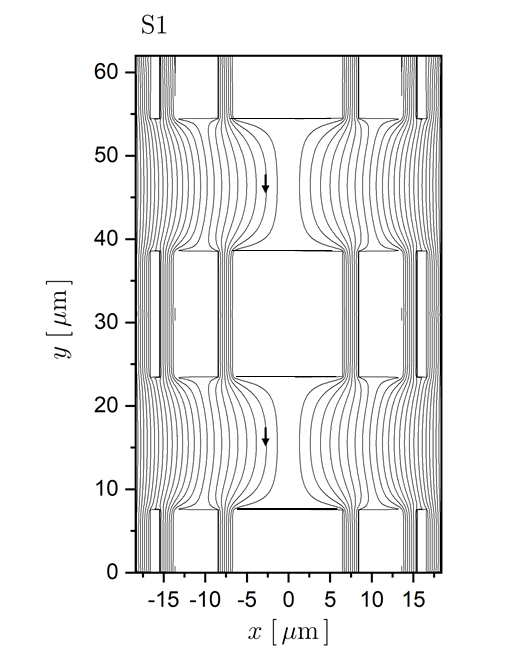}
\label{default}
\hspace*{-8mm}
\includegraphics[width=0.55\textwidth]{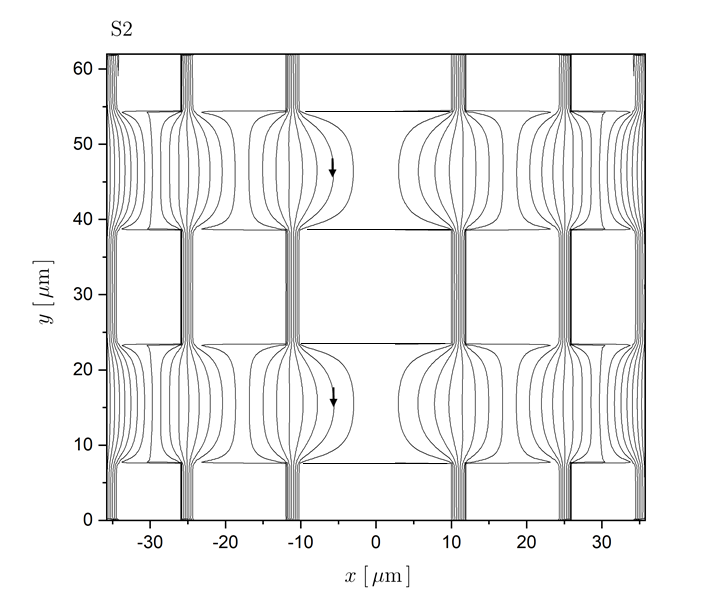}
\caption{Current streamlines of $\vec{j}^{(I_b)}$ in S1 and S2 to determine $\vec{L}^{I_b}$.}
\label{default}
\end{center}
\end{figure}

\begin{figure}[! h]
\begin{center}
\hspace*{-5mm}
\includegraphics[width=0.50\textwidth]{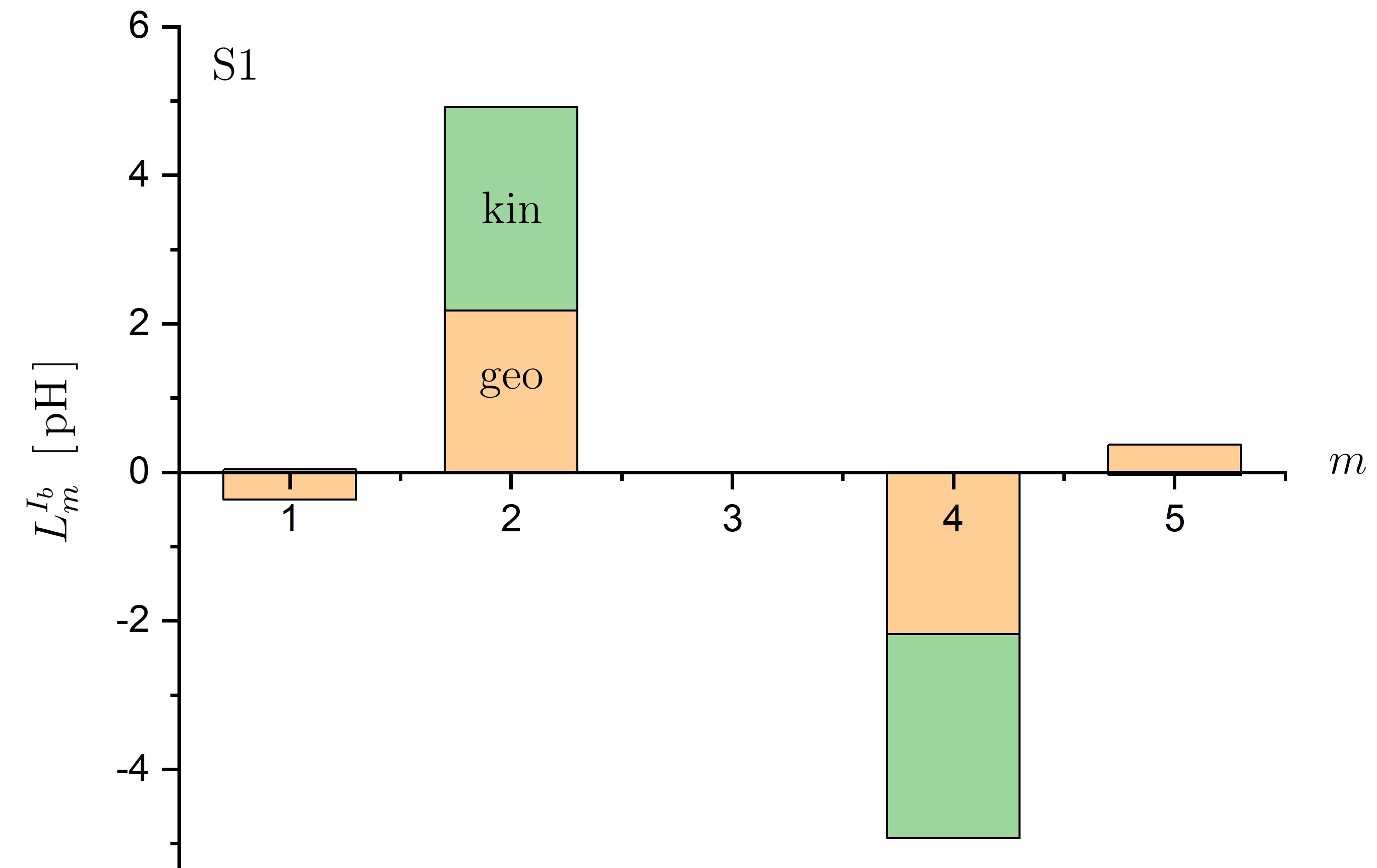}
\label{default}
\hspace*{-5mm}
\includegraphics[width=0.50\textwidth]{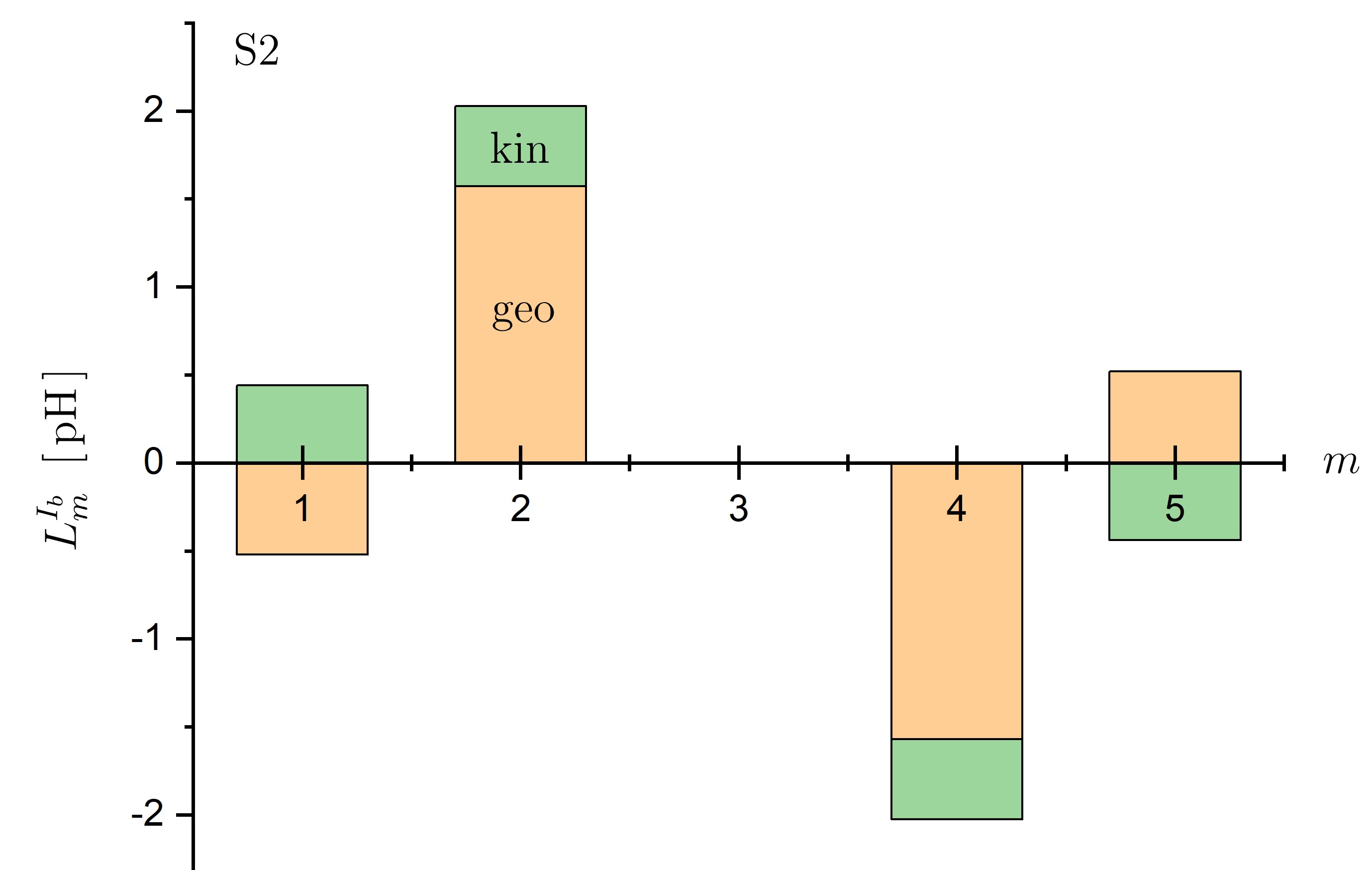}
\caption{Components $L_m^{I_b}$ of the bias current inductance vector $\vec{L}^{I_b}$. The geometric and kinetic parts are shown.}
\label{default}
\end{center}
\end{figure}

Figure 19 shows $L_m^{I_b}$ for the different holes $m$ along a row, and the geometric and kinetic parts are indicated.
 \\
 \par
 \section{Reflection asymmetry of $V(B_a)$}

\begin{figure}[! h]
\begin{center}
\hspace*{-7mm}
\includegraphics[width=0.5\textwidth]{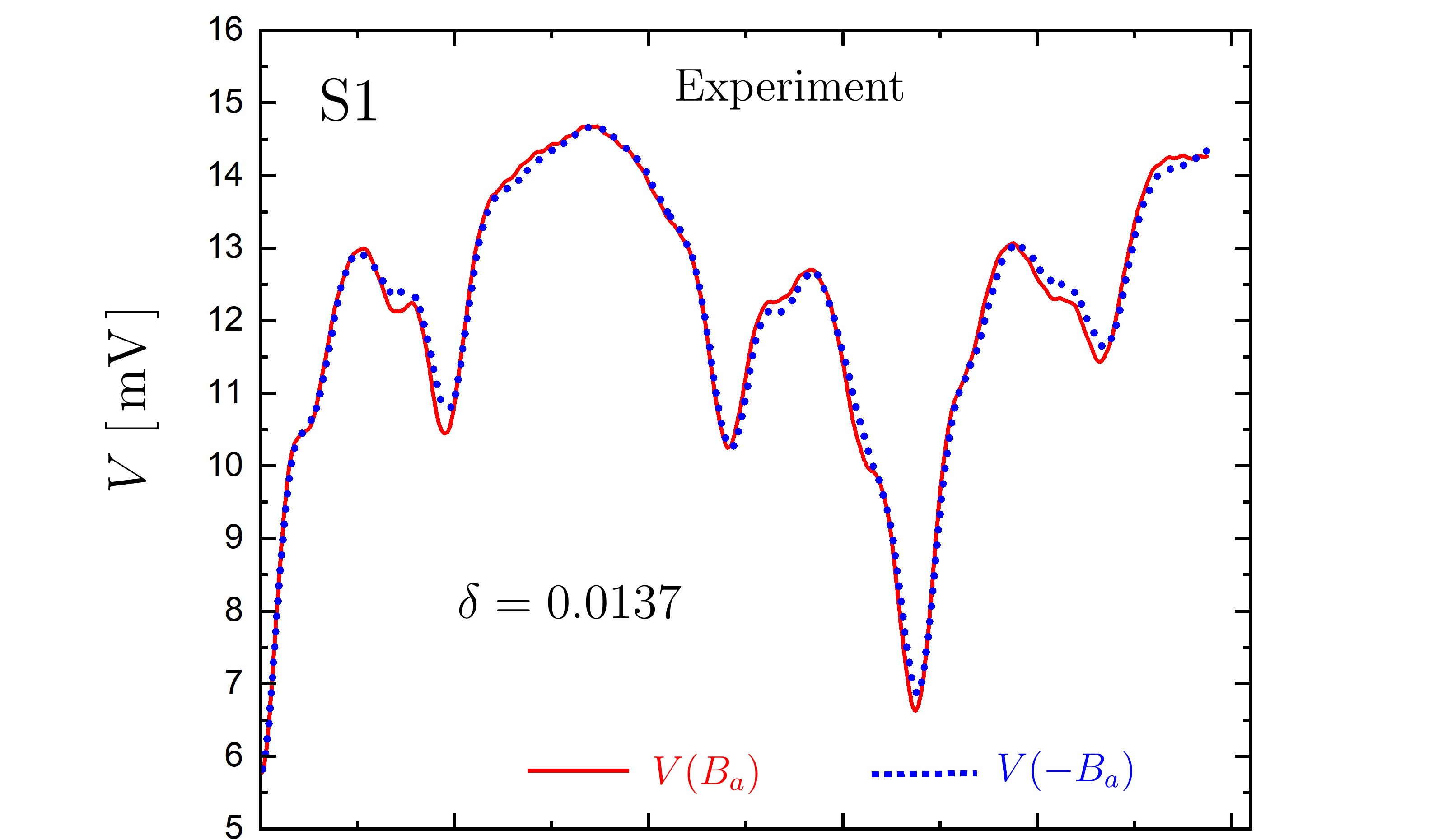}
\label{default}
\hspace*{-7mm}
\includegraphics[width=0.5\textwidth]{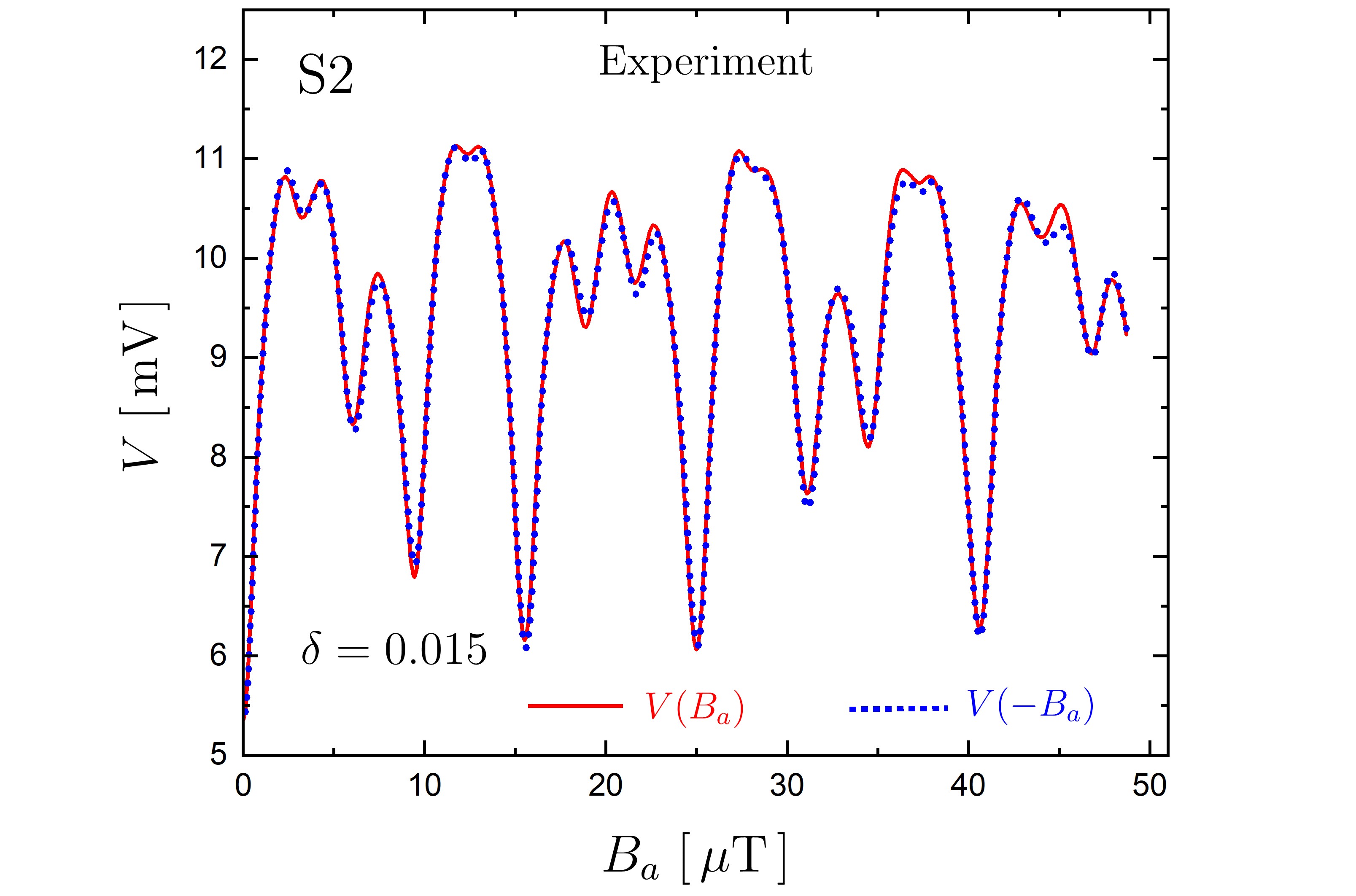}
\caption{Experimental $V(B_a)$ and $V(-B_a)$ of array S1 and S2 to show the reflection asymmetry.}
\label{default}
\end{center}
\end{figure}

\begin{figure}[!h]
\begin{center}
\hspace*{-7mm}
\includegraphics[width=0.5\textwidth]{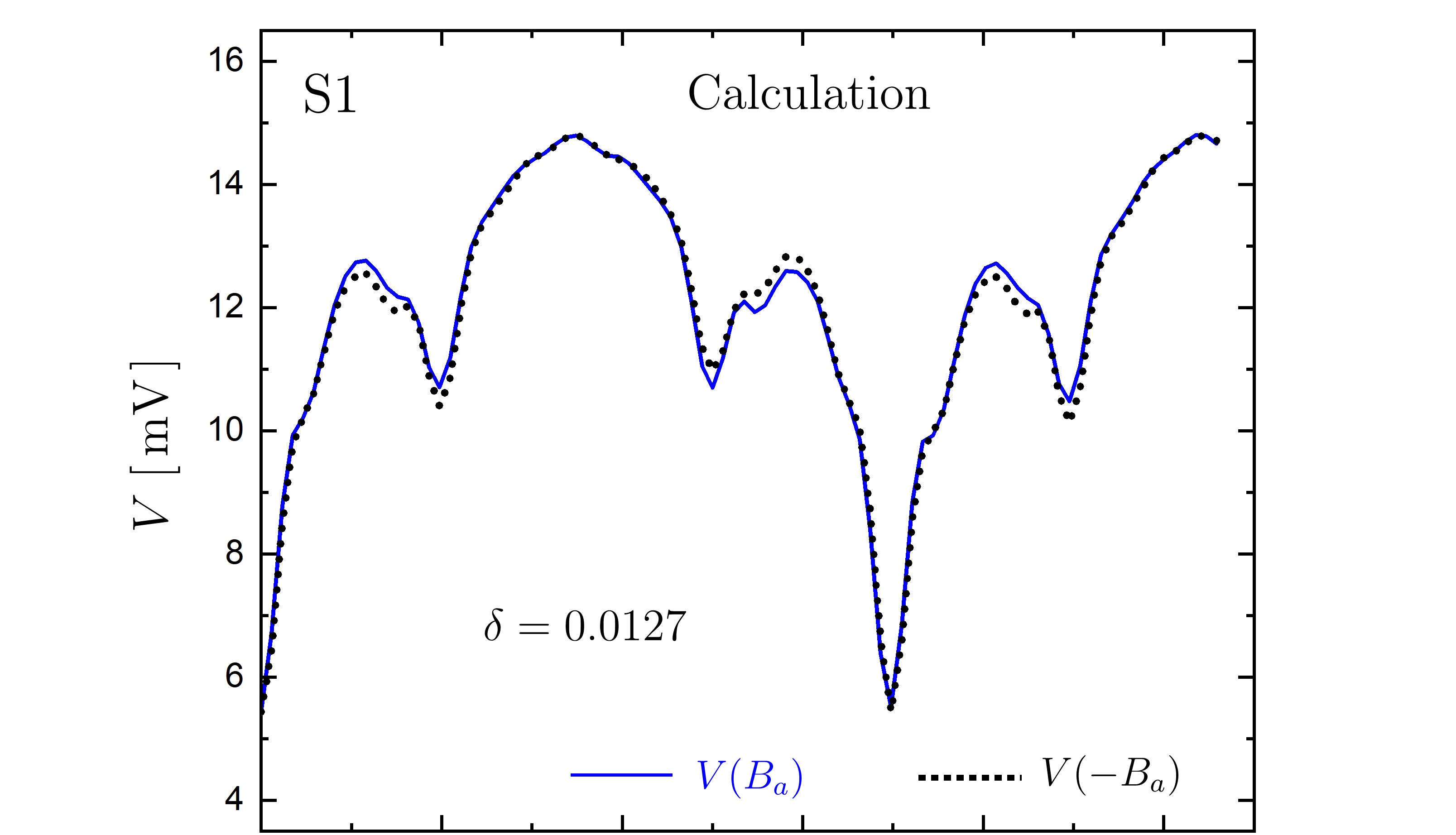}
\label{default}
\hspace*{-7mm}
\includegraphics[width=0.5\textwidth]{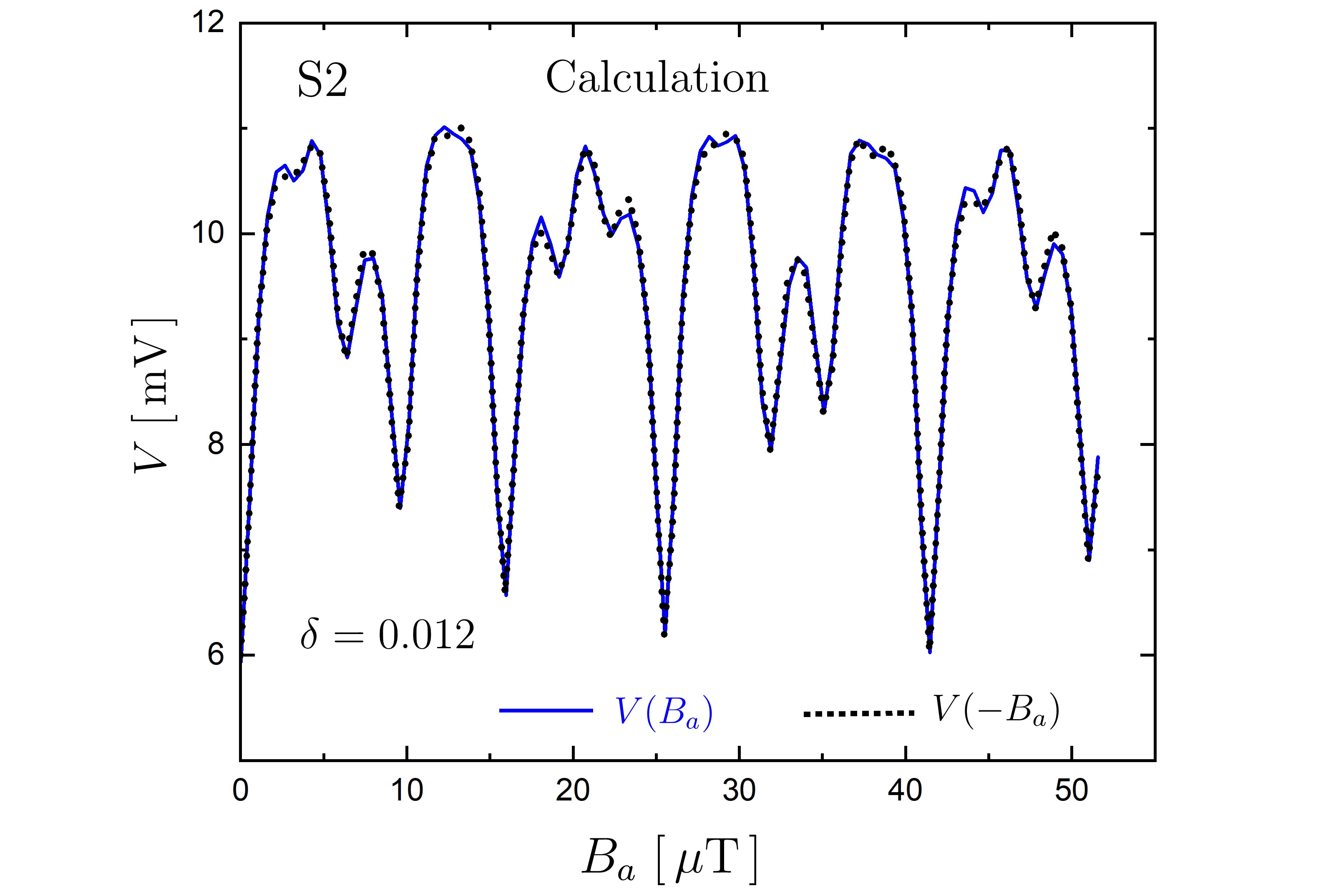}
\caption{Calculated $V(B_a)$ and $V(-B_a)$ of array S1 and S2 to show the reflection asymmetry.}
\label{default}
\end{center}
\end{figure}

Figure 20 shows the experimental reflection asymmetry of $V(B_a)$ by plotting $V(B_a)$ and $V(-B_a)$ for S1 and S2. Small differences between $V(B_a)$ and the reflected curve $V(-B_a)$ are noticeable. Figure 21 displays the calculated $V(B_a)$ and $V(-B_a)$ curves. One can define the degree $\delta$ of the reflection asymmetry as
\begin{equation}
\delta = \frac{\int_{-\Delta B / 2}^{\Delta B / 2} \, \left | \; V(B_a) - V(-B_a) \; \right | \, dB_a}{ \Delta V \Delta B} \;  ,
\end{equation}
with the integration interval $\Delta B = 100 \, \mu$T. 
The experimental $\delta$ for S1 and S2 are $\delta = 0.0137$ and 0.015, respectively, while the calculated ones are $\delta = 0.0127$ and 0.012.
We found from our calculations that $\delta$ increases when the $I_c$-spread $\sigma$ increases.
\clearpage
\bibliography{JL542}

\end{document}